\documentclass[aps,prb,twocolumn,showpacs,superscriptaddress,longbibliography]{revtex4-2}
\usepackage{amsfonts}
\usepackage{amsmath}
\usepackage{graphicx}
\usepackage{bm}
\usepackage{amssymb}
\usepackage{dcolumn}
\usepackage{color}
\usepackage{multirow}
\usepackage{booktabs}
\usepackage[colorlinks,
linkcolor=blue,
anchorcolor=blue,
citecolor=blue,
urlcolor=blue,
]{hyperref}
\begin{document}	
		\title{Superconducting, topological and transport properties of kagome metals CsTi$ _{3} $Bi$ _{5} $ and RbTi$ _{3} $Bi$ _{5} $}
\begin{titlepage}
	\vspace*{\stretch{1}}

	\begin{center}
		{\huge\bfseries Superconducting, topological and transport properties of kagome metals CsTi$ _{3} $Bi$ _{5} $ and RbTi$ _{3} $Bi$ _{5} $}                  \\[6.5ex]
		{\large\bfseries Xin-Wei Yi,$ ^{1} $  Zheng-Wei Liao,$ ^{1} $  Jing-Yang You,$ ^{2, *} $  Bo Gu,$ ^{1,3, \dagger} $  and Gang Su $ ^{1,3, \ddagger} $ } \\
		\vspace{4ex}
		{\normalsize \bfseries $ 1 $  School of Physical Sciences, University of Chinese Academy of Sciences, Beijing 100049, China}\\[2ex]
		{\normalsize \bfseries $ 2 $  Department of Physics, Faculty of Science, National University of Singapore, 117551, Singapore}\\[2ex]
		{\normalsize \bfseries $ 3 $  Kavli Institute for Theoretical Sciences, and CAS Center for Excellence in Topological Quantum Computation, University of Chinese Academy of Sciences, Beijing 100190, China}\\[2ex]
		{\normalsize \bfseries $ *,\dagger,\ddagger $  Corresponding authors}\\[1ex]		
		{\normalsize \bfseries  Email addresses:\\ phyjyy@nus.edu.sg (Jing-Yang You), gubo@ucas.ac.cn (Bo Gu), gsu@ucas.ac.cn (Gang Su)}\\[2ex]	
		\vfill
		June 26, 2023
	\end{center}
	\vspace{\stretch{2}}

	\newpage
	
\end{titlepage}

	\begin{abstract}
\textbf{Abstract:} The recently discovered ATi$_3$Bi$_5$ (A=Cs, Rb) exhibit intriguing quantum phenomena including superconductivity, electronic nematicity, and abundant topological states, which provide promising platforms for studying kagome superconductivity, band topology, and charge orders. In this work, we comprehensively study various properties of ATi$_3$Bi$_5$ including superconductivity under pressure and doping, band topology under pressure, thermal conductivity, heat capacity, electrical resistance, and spin Hall conductivity (SHC) using first-principles calculations. Calculated superconducting transition temperature ($\mathrm{ T_{c}}$) of CsTi$_3$Bi$_5$ and RbTi$_3$Bi$_5$ at ambient pressure are about 1.85 and 1.92K. When subject to pressure, $\mathrm{ T_{c}}$ of CsTi$_3$Bi$_5$ exhibits a special valley and dome shape, which arises from quasi-two-dimensional to three-dimensional isotropic compression within the context of an overall decreasing trend. Furthermore, $\mathrm{ T_{c}}$ of RbTi$_3$Bi$_5$ can be effectively enhanced up to 3.09K by tuning the kagome van Hove singularities (VHSs) and flat band through doping. Pressure can also induce abundant topological surface states at the Fermi energy ($\mathrm{E}_{\mathrm{F}}$) and tune VHSs across $\mathrm{E}_{\mathrm{F}}$. Additionally, our transport calculations are in excellent agreement with recent experiments, confirming the absence of charge density wave. Notably, SHC of CsTi$_3$Bi$_5$ can reach as large as 226$ \hbar $·(e·$ \Omega $·cm)$ ^{-1} $ at $\mathrm{E}_{\mathrm{F}}$. Our work provides a timely and detailed analysis of the rich physical properties for ATi$_3$Bi$_5$, offering valuable insights for further explorations and understandings on these intriguing superconducting materials.

\begin{description}
	\item[keywords]
	CsTi$_3$Bi$_5$; RbTi$_3$Bi$_5$; Superconductivity; Electronic topology; Transport properties; Spin\\ Hall effect.
\end{description}
	\end{abstract}

	\maketitle

\section{INTRODUCTION}
The kagome lattice composed of tiled corner-sharing triangles is a well-known two-dimensional (2D) prototype lattice, in which topological magnetism, superconductivity and frustration have been extensively studied in recent decades \cite{RN1235, RN976}. The ideal electronic band structure of kagome lattice is characterized by flat bands, Dirac cones, quadratic band touching points and van Hove singularities (VHSs) (Fig. S1(b)). The flat bands originating from the destructive interference of wavefunctions suggest a significant correlation between electrons and can induce special negative magnetism \cite{RN100}, fractional quantum Hall effect \cite{RN109, RN110}, Wigner crystallization \cite{RN116}, high-temperature superconductivity \cite{RN1231}, etc. Moreover, the spin-orbit coupling (SOC) at the massless Dirac points and quadratic band touching points introduces energy gaps, which can give rise to nontrivial Z$ _{2} $ topology akin to topological insulators. These gaps can even exhibit Chern gaps when subjected to magnetization \cite{RN1229, RN1230}. Fermi surface instabilities due to sublattice interference near VHS may cause different long-range charge orders and unconventional superconductivity \cite{RN16, RN856, RN19}. Additionally, other novel physical properties, including Wely points \cite{RN104, RN502, RN1253}, giant anomalous Hall effect \cite{RN99, RN264, RN92}, quantum anomalous Hall effect \cite{RN108, RN232}, topological superconductors \cite{RN253, RN514}, are all intriguing research topics of kagome structure.

The recently discovered AV$_3$Sb$_5$ (A=K, Rb, Cs) kagome family exhibits various attractive properties \cite{RN5}. As the pioneering examples of quasi-2D kagome superconductors, their superconducting transition temperatures ($\mathrm{ T_{c}}$) are about 1.8-2.5K under ambient pressure \cite{RN11, RN8, RN64, RN34}. Alongside the emergence of 2×2 charge density wave (CDW) induced by VHSs with transition temperatures of about 100K \cite{RN132, RN52, RN9}, AV$_3$Sb$_5$ hosts a variety of unconventional charge orders, including a nematic phase reminiscent of iron-based superconductors \cite{RN187, RN926}, a pair density wave similar to copper-based superconductors \cite{RN34}, and a 4×1 CDW that induces striking phase fluctuation \cite{RN222, RN1216}. Pressure and doping are two effective ways to manipulate the phase diagram of AV$_3$Sb$_5$. Multiple superconductivity domes simultaneously emerge intertwined with several CDW orders under pressure \cite{RN926, RN469, RN9, RN54}. Experimental substitution doping with Ti, Nb, Ta, Mo, Mn, Cr, Sn, and As in AV$_3$Sb$_5$ have been achieved, providing control over superconductivity, CDW, and nematicity \cite{RN1203, RN1158, RN949, RN204, RN1241, RN204, RN1239, RN1014, RN1240, RN1238, RN202}. CsV$ _{3} $Sb$ _{5} $, in particular, possesses both abundant non-trivial Z$_2$ topological surface states (TSS) and superconductivity \cite{RN8, RN461}, and has been instrumental in resolving possible Majorana zero modes within vortex cores \cite{RN1}.

High-throughput density functional theory (DFT) calculations have been employed to analyze a range of compounds based on the AV$_3$Sb$_5$ prototype structure, revealing 24 dynamically stable compounds with superconductivity, abundant TSS and CDW \cite{RN985}. Among them, ATi$_3$Bi$_5$ (A=Rb, Cs) compounds with enhanced $\mathrm{ T_{c}}$ of about 4.8K and electronic nematicity have recently been synthesized \cite{RN565, RN967, RN795, RN955}. Another kagome superconductor family, similar to ATi$_3$Bi$ _{5} $, has also been predicted and awaits experimental verification \cite{RN1006}. The nematic order in CsV$_3$Sb$_5$ is considered as the vestige order of CDW \cite{RN187}. However, the original lattice structure of ATi$_3$Bi$_5$ has been confirmed to be highly stable and devoid of CDW, as supported by calculations and thermal transport measurements \cite{RN985, RN565, RN967, RN1226}. The presence of orbital-selective nematic order and intertwined superconductivity in ATi$_3$Bi$_5$ provides a new platform for exploring the multi-orbital correlation effect in nematic superconductors. In addition, exotic electronic structures have been observed in ATi$_3$Bi$_5$ using angle-resolved photoemission spectroscopy (ARPES), including flat band, type II and type III Dirac nodal lines and Z$_2$ TSS \cite{RN966, RN970, RN968, RN1018, RN965, RN1123}. These fascinating properties make ATi$_3$Bi$_5$ as ideal systems for investigating various kagome-related physics with reference to AV$_3$Sb$_5$. Consequently, further investigations are imperative, particularly in examining the effects of pressure and doping on the properties of superconductivity and topology.

In this paper, we explore the superconducting and topological properties of ATi$_3$Bi$_5$ under pressure and doping by first-principles calculations. Estimated $\mathrm{ T_{c}}$ of CsTi$_3$Bi$_5$ and RbTi$_3$Bi$_5$ at ambient pressure are about 1.85 and 1.92K. We observe the emergence of a special valley and dome in the $\mathrm{ T_{c}}$ of CsTi$_3$Bi$_5$ under pressure, which is due to the crossover from quasi-2D to three-dimensional (3D) isotropic compression within the range of 10-20GPa, accompanied by a decreased background effect across all pressure ranges. Moreover, electron-phonon coupling (EPC) calculations with both a rigid band model and atomic substitution doping confirm the substantial enhancement of superconductivity in RbTi$_3$Bi$_5$ under doping. Additionally, pressure can lead to the presence of abundant TSS near $\mathrm{E}_{\mathrm{F}}$. We also discuss properties such as thermal conductivity, heat capacity, and electrical resistance, which are consistent with recent experimental findings, indicating the absence of CDW. We find that CsTi$_3$Bi$_5$ exhibits an intrinsic SHC of as large as 226$ \hbar $·(e·$ \Omega $·cm)$ ^{-1} $ at the Fermi energy ($\mathrm{E}_{\mathrm{F}}$). Our findings indicate that both doping and pressure are effective means of tuning the electronic band structures and supercondutvitity of kagome ATi$_3$Bi$_5$.

\section{Results of C$\lowercase{\textbf{s}}$T$\lowercase{\textbf{i}}$$ _{3} $B$\lowercase{\textbf{i}}$$ _{5}$}
\subsection{Electronic Structure and van Hove Singularities}	

\begin{figure}[t]
	\centering
	\includegraphics[scale=0.51,angle=0]{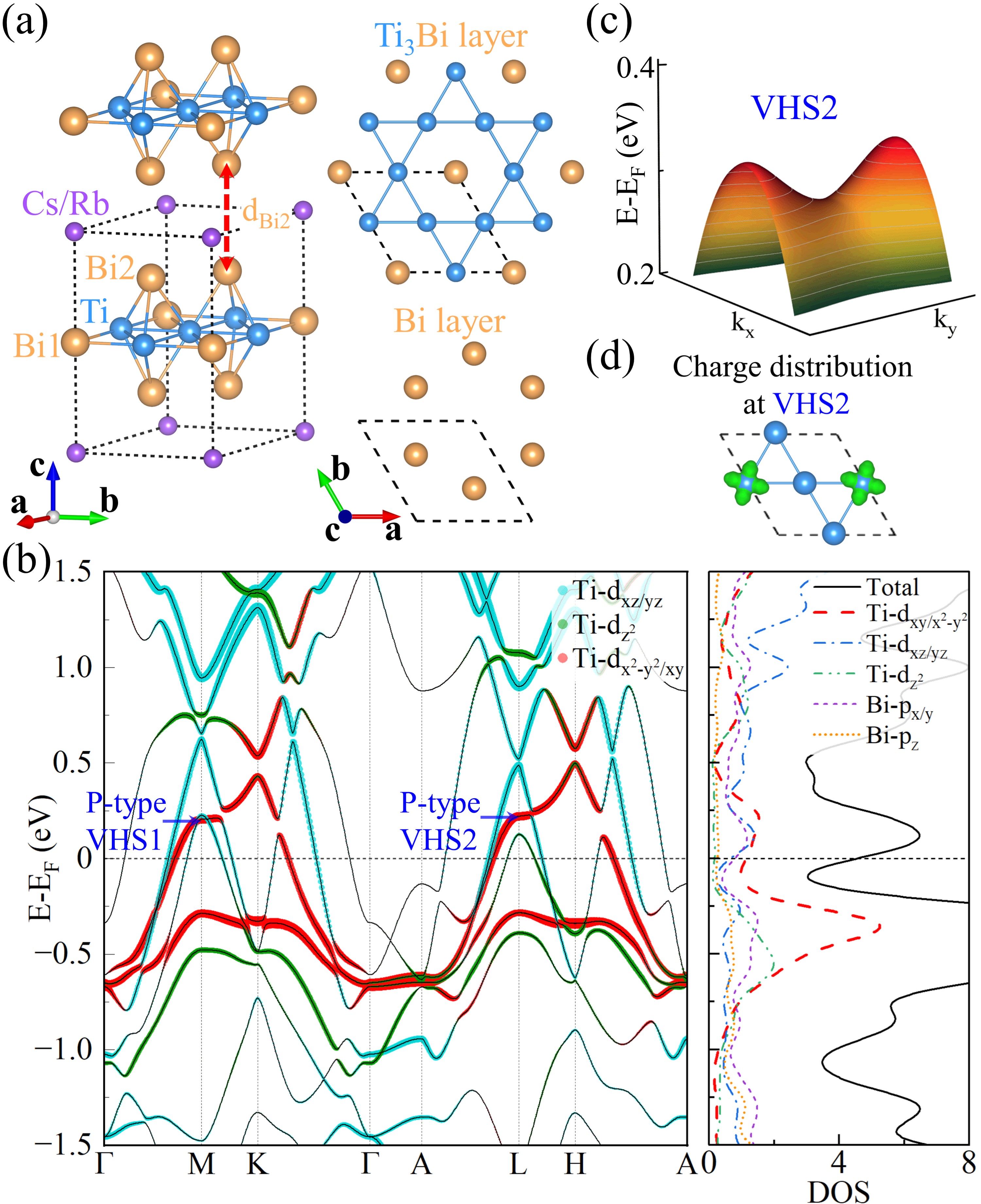}\\
	\caption{\textbf{The crystal structure and electronic band structure of CsTi$_\mathbf{3}$Bi$_\mathbf{5}$.} (a) The side view of crystal structure and top views of Ti$ _{3} $Bi kagome layer and Bi hexagonal layer. “d$ _{\mathrm{Bi2}} $” represents the vertical distance between Bi2 atoms in the adjacent primitive lattices, indicated by the black dotted lines. (b) Calculated electronic band structure projected by different orbitals of Ti atoms and partial density of states (DOS) with spin-orbit coupling (SOC) for CsTi$_3$Bi$_5$. Two kagome VHSs near the Fermi level are labeled as “VHS1” and “VHS2”, respectively. (c) 3D band structure near VHS2, with gray curves indicating constant energy contours. (d) Charge density distribution of the electronic state at VHS2 (isovalue: 0.003 e/bohr$^3$).}\label{band}
\end{figure}
The ATi$_3$Bi$_5$ compounds exhibit a layered structure with the space group P6/mmm (No. 191) as illustrated in Fig.~\ref{band}(a). The central Ti$ _{3} $Bi layer consists of a Ti-kagome net intertwined with a Bi triangular net, sandwiched between two Bi honeycomb layers. The triangular layers composed of A atoms are loosely bonded with the central Ti-Bi sheets. Bi atoms occupy two distinct Wyckoff positions, designated as Bi1 and Bi2, respectively. The calculated electronic band structures and partial electronic density of states (DOS) with SOC for CsTi$_3$Bi$_5$ are plotted in Fig.~\ref{band}(b). By assigning colors to the bands based on the weights of different Ti atomic orbitals, we can identify a clear kagome flat band at about -0.4eV, VHSs at M and L points (namely VHS1 and VHS2) at about 0.2eV and massive Dirac points at K and H points contributed by Ti-d$ _{x^{2}-y^{2}/xy} $ orbitals. The VHSs and flat band of Ti-d$ _{x^{2}-y^{2}/xy} $ orbitals give rise to two DOS peaks at the corresponding energy levels. Notably, clear saddle features are observed in the 3D band structure near VHS2 as plotted in Fig.~\ref{band}(c).

 Two distinct VHSs arise from the general kagome electronic bands, namely p-type and m-type VHSs, corresponding to pure and mixed sublattice characteristics at the VHS, respectively. The charge density distribution of the electronic state at the VHS2 shows pronounced in-plane d-orbital features and a p-type character as illustrated in Fig.~\ref{band}(d). Unlike CsV$_3$Sb$_5$, where all p-type VHSs are located above corresponding Dirac points, and the p-type VHS1 and VHS2 of CsTi$_3$Bi$_5$ lie below Dirac points in Fig.~\ref{band}(b). To clarify this, we construct a tight-binding (TB) kagome model with a single orbital on each sublattice and nearest-neighboring (NN) hopping and a three-band Hamiltonian can be written as follows
	\begin{equation}
		H=\sum_{\substack{i, j \\ i \neq j}} 2 t \cos (k \cdot d_{i j}) c_{i,k}^{\dagger } c_{j,k}, \quad (i, j=A, B, C)
	\end{equation}
	where $ \mathit{t} $, $d_{ij}$ and $k=(k_x,k_y,k_z )$ represent the NN hopping parameter, NN space vector of sublattice $i$ and $j$, and wavevector of reciprocal space, respectively. $ \mathit{A} $, $ \mathit{B} $, and $ \mathit{C} $ represent three kagome sublattices. This toy kagome model produces the flat band spanning the entire Brillouin zone (BZ), VHSs at $ \bar{\mathrm{M}} $, Dirac point at $ \bar{\mathrm{K}} $, and quadratic band touching point at $ \bar{\Gamma} $ as shown in Fig. S1(b) of the supplemental information (SI). By adjusting $ \mathit{t} $, it becomes apparent that for $ \mathit{t} $$>$0, the p-type and m-type VHSs are located above and below the Dirac points, respectively, and the flat band resides at the top of band structure. Conversely, for t$ < $0, the bands are reversed. This observation indicates that in the kagome sublattice of CsV$_3$Sb$_5$, all the relevant NN hopping parameters are positive, while for the Ti-d$ _{x^{2}-y^{2}/xy} $ orbitals in CsTi$_3$Bi$_5$, the NN hopping parameters exhibit negative values. P-type VHSs can induce Fermi surface instabilities \cite{ RN16, RN856, RN19}, and further investigations are warranted to explore their connection with the nematic and superconducting properties in CsTi$_3$Bi$_5$.
	
\subsection{Transport properties}
\begin{figure}[t]
	\centering
	\includegraphics[scale=0.45,angle=0]{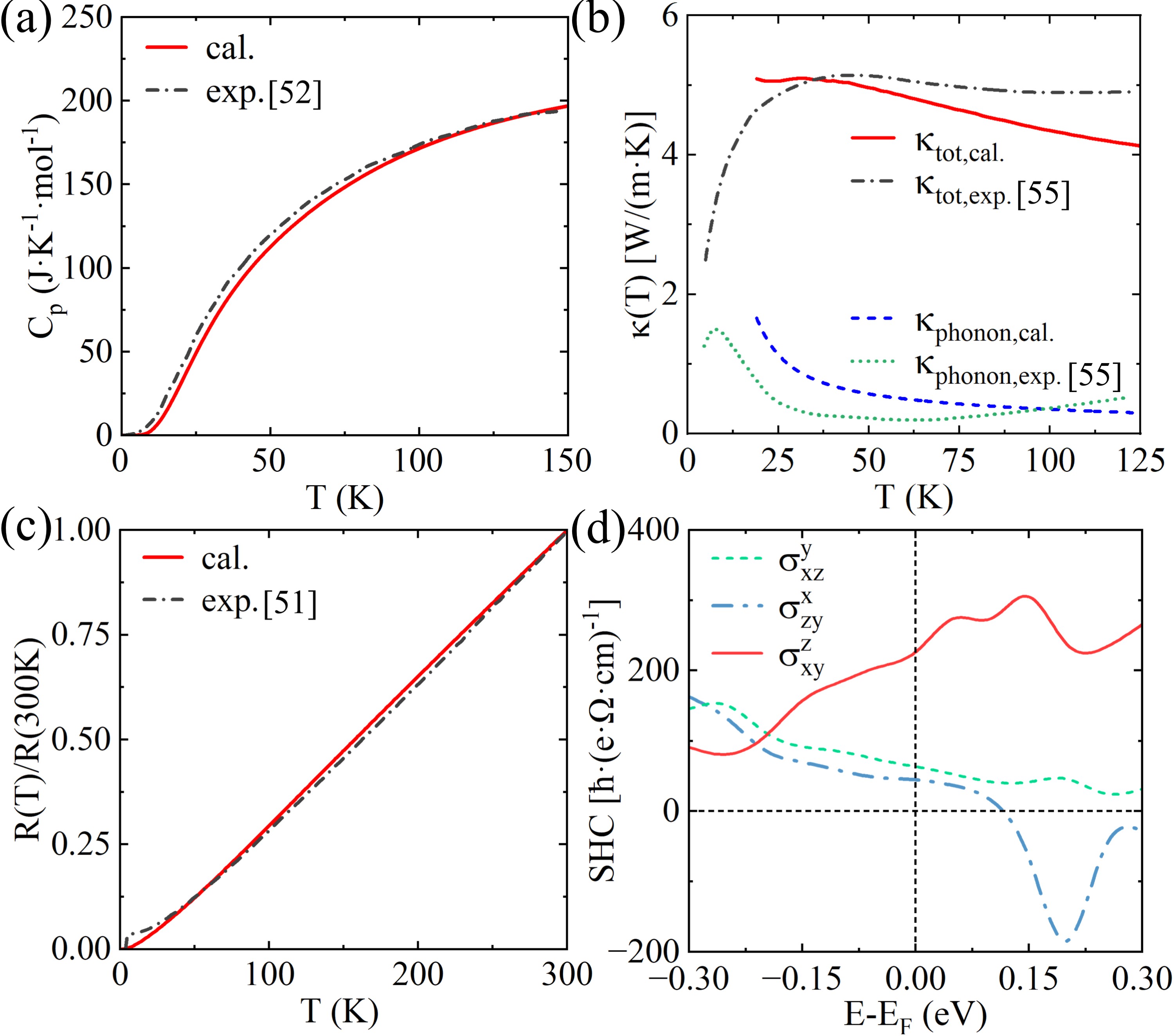}\\
	\caption{\textbf{Heat capacity and transport properties of CsTi$_\mathbf{3}$Bi$_\mathbf{5}$.} (a) Calculated temperature dependence of the heat capacity compared with the experiment \cite{RN795}. (b) Calculated temperature dependence of the longitudinal thermal conductivity compared with the experiment \cite{RN1226}. (c) Calculated temperature dependence of the longitudinal electrical resistivity compared with the experiment \cite{RN967}. (d) Calculated three independent components of spin Hall conductivity (SHC) tensor as a function of energy.}\label{trans}
\end{figure}

To gain insight into the transport properties of ATi$_3$Bi$_5$, we calculate the temperature dependence of heat capacity C$_p$, thermal conductivity $\kappa$ and resistance R for CsTi$_3$Bi$_5$. These results are compared with recent experimental data as plotted in Figs. \ref{trans}(a-c). The calculated results for RbTi$_3$Bi$_5$ are listed in Fig. S2. Overall, our transport calculations are in quantitative agreement with the experimental findings, demonstrating the reliability of our method. Obviously, there are no phase transition signatures in any of the transport quantities, suggesting the absence of CDW. In contrast to CsV$_3$Sb$_5$, the total thermal conductivity $\kappa_{\mathrm{tot, cal}}$ of CsTi$_3$Bi$_5$ is primarily governed by electronic contributions rather than phonon thermal conductivity $\kappa_{\mathrm{phonon, cal}}$ \cite{RN1226}. We choose the results with the highest residual resistance ratios from Ref. \cite{RN967}, where the impurity concentration is the lowest in that experiment. The resistance of CsTi$_3$Bi$_5$, as shown in Fig. S2(b), is much smaller than that of AV$_3$Sb$_5$ \cite{RN8}. The calculated R(T)/R(300K) values align well with the experimental results at high temperatures, but underestimate the experimental data below 30K as shown in Fig. \ref{trans}(c). This discrepancy arises that our resistance calculations solely consider the electron-phonon interaction and do not account for electron-electron interactions. In the experimental resistance-temperature behavior, a quadratic dependence below 30K and a linear dependence above 50K can be observed, corresponding to electron-electron and electron-phonon interactions in the normal metal, respectively. It is important to note that our calculations of the transport properties are performed on the normal phase of CsTi$_3$Bi$_5$, and not on the superconducting phase.

In addition to the aforementioned transport properties, we also calculate the three independent components of SHC tensor, i.e. $ \sigma_{xy}^{z} $, $ \sigma_{zx}^{y} $, and $ \sigma_{xz}^{y} $, as a function of the chemical potential for CsTi$_3$Bi$_5$ in Fig.\ref{trans}(d). The magnitudes of three components can reach as large as 226$ \hbar $·(e·$ \Omega $·cm)$ ^{-1} $ at $\mathrm{E}_{\mathrm{F}}$. The large SHC comes from narrow band gap opened by large SOC \cite{RN1206}. The large SHC observed in CsTi$_3$Bi$_5$ implies great potential for applications in spintronic devices.

\subsection{Supercondutivity under Pressure}

\begin{figure*} [t]
	\centering
	\includegraphics[scale=0.477,angle=0]{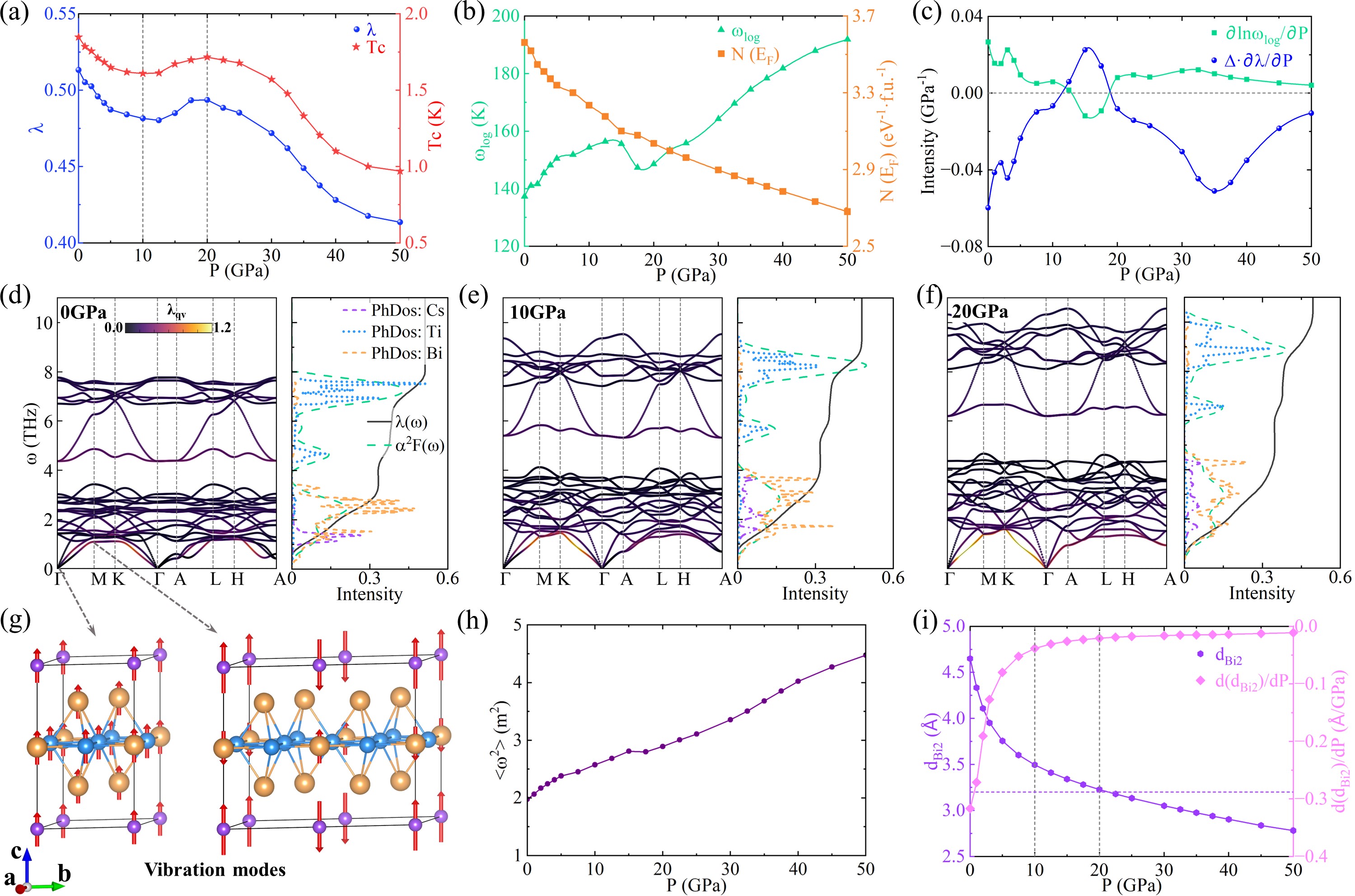}\\
	\caption{\textbf{Superconducting properties of CsTi$_\mathbf{3}$Bi$_\mathbf{5}$ under pressure in the range of 0-50GPa.} (a) Superconducting transition temperature $\mathrm{ T_{c}}$ and electron-phonon coupling (EPC) $ \lambda $ under pressure. (b) Logarithmic average frequency $\omega_ {log} \mathrm{(K)}$ and electronic DOS per formula unit (f.u.) at the Fermi energy N(E$_F$) under pressure. (c) $ \partial \ln \omega_{\log } / \partial \mathrm{P} $ and $\Delta\cdot\partial \lambda / \partial \mathrm{P}$ under pressure. Phonon spectra colored by EPC strength $\lambda_{\mathbf{q} \nu}$, projected phDOS, Eliashberg spectral function $\alpha^2F(\omega)$, and cumulative frequency-dependent EPC $ \lambda(\omega) $ at (d) 0, (e) 10, and (f) 20GPa, respectively. (g) Vibration modes at the lowest acoustic branch near $ \Gamma $ and M points, both of which are contributed by the vibrations along the c axis. (h) The average of square of phonon frequency $\left\langle\omega^2\right\rangle$ under pressure. (i) d$ _{\mathrm{Bi2}} $ denoted in Fig.~\ref{band}(a) and d(d$ _{\mathrm{Bi2}} $)/dP under pressure. The horizontal dashed line indicates the Bi atomic diameter of 3.2\AA.}\label{presssup}
\end{figure*}

The superconducting $\mathrm{ T_{c}}$ of CsV$ _{3} $Sb$ _{5} $ at pressures above 2GPa, in the absence of CDW, have been found in good agreement between EPC calculations and experimental results \cite{RN150, RN193}. Therefore, we can also expect DFT calculations to accurately capture the true superconductivity in ATi$_3$Bi$_5$. The estimated $\mathrm{ T_{c}}$'s of CsTi$_3$Bi$_5$ and RbTi$_3$Bi$_5$ at ambient pressure are about 1.85 and 1.92K, respectively. The superconducting properties of both materials share remarkable similarities, so we choose CsTi$_3$Bi$_5$ as an instance to study the superconductivity under pressure.

Our calculated $\mathrm{ T_{c}}$ and EPC constant $\lambda$ as a function of pressure are shown in Fig.~\ref{presssup}(a). Both values decrease in the ranges of 0-10 and 20-50 GPa, while they increase within the range of 10-20GPa, forming a unique valley and dome shape under pressure. The corresponding logarithmic average of the phonon frequency $\omega_{log}$ and the electronic DOS at the Fermi energy N$(\mathrm{E}_{\mathrm{F}})$ are shown in Fig.~\ref{presssup}(b). N$(\mathrm{E}_{\mathrm{F}})$ shows an almost linear decrease with increasing pressure, while $\omega_{log}$ exhibits an overall increase but with a decline specifically within the 10-20GPa. It is noted that $\mathrm{ T_{c}}$ is explicitly proportional to $\omega_{log}$ as described by McMillan semi-empirical formula \cite{RN128, RN129}, ${\rm T_c}=\frac{\omega_{log}}{1.2}{\rm exp}[{-\frac{1.04(1+\lambda)}{\lambda-\mu^*(1+0.62\lambda)}}]$, whereas the variation of $\mathrm{ T_{c}}$ in CsTi$_3$Bi$_5$ shows an opposite trend compared with $\omega_{log}$ in Figs.~\ref{presssup}(a) and (b). To address this apparent contradiction, we differentiate McMillan formula with respect to P, which gives

\begin{equation}
	\begin{aligned}
		&\frac{d T_c}{d P}=T_c \cdot\left[\partial \ln \omega_{\log }/\partial P+\Delta\cdot\partial \lambda / \partial P\right],\\ \\
		\mathrm{and} \quad &\Delta=\frac{1.04\left(1+0.38 \mu^*\right)}{\left[\lambda-\mu^*(1+0.62 \lambda)\right]^2}.
	\end{aligned}
\end{equation}

We calculate two terms in the brackets separately and illustrate them in Fig.~\ref{presssup}(c). It becomes evident that $\lambda$ and $\omega_{log}$ make totally opposite contributions to $\frac{d T_c}{d P}$ across all pressure ranges. However, $\partial \ln \omega_{\log } / \partial \mathrm{P} $ is much smaller than $\Delta \cdot \partial \lambda / \partial \mathrm{P}$. This observation explains the reason why $\mathrm{ T_{c}}$ follows the same trend as $\lambda$ but not $\omega_{log}$ under pressure.

The remaining question is why $\lambda$ ($\mathrm{ T_{c}}$) displays a unique valley and dome shape under pressure. To shed light on this, we present the phonon spectra, projected phonon DOS (phDOS), Eliashberg spectral function $\alpha^2F(\omega)$, and cumulative frequency-dependent EPC $ \lambda(\omega) $ in Figs.~\ref{presssup}(d-f). The corresponding results under other pressures can be found in Fig. S3. As pressure increases, the phonon frequencies ($\omega$) within the phonon spectra noticeably increase, which is a common characteristic observed in many materials. This makes the average of the square of phonon frequency $\left\langle\omega^2\right\rangle$ almost linearly increase under pressure as seen in Fig.~\ref{presssup}(h). From the definition of $\lambda$, we have $\lambda=\frac{N\left(\mathrm{E}_{\mathrm{F}}\right)\left\langle I^2\right\rangle}{M\left\langle\omega^2\right\rangle}$ and 
	$\left\langle\omega^2\right\rangle=\left[\frac{2}{\lambda} \int_0^{\infty} \alpha^2 F(\omega) \omega d \omega\right]$,
where $\left\langle I^2\right\rangle$ is the mean-square electron-ion matrix element, and M is the ionic mass. It is clear that $\lambda$ is proportional to N$(\mathrm{E}_{\mathrm{F}})$ and inversely proportional to $\left\langle\omega^2\right\rangle$. The combination of N$(\mathrm{E}_{\mathrm{F}})$ and $\left\langle\omega^2\right\rangle$ results in an overall decrease in $\lambda$. This mechanism accounts for the downward trend of $\lambda$ and $\mathrm{ T_{c}}$ within the range of 0-10 and 20-50GPa.

\begin{figure*}[t]
	\centering
	\includegraphics[scale=0.42,angle=0]{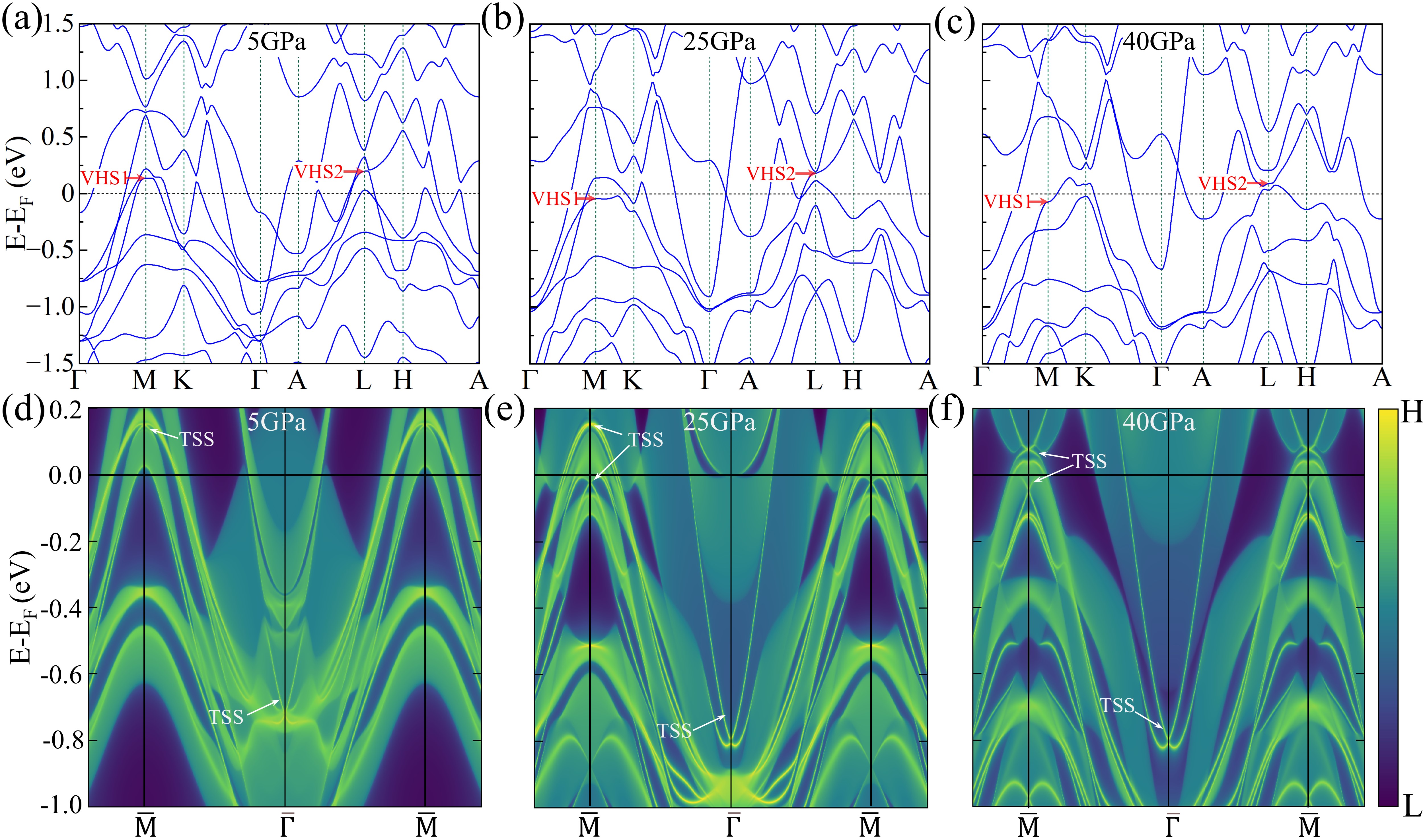}\\
	\caption{\textbf{Electronic band structures and surface state spectra under pressure for CsTi$_\mathbf{3}$Bi$_\mathbf{5}$.} electronic band structures with SOC at (a) 5, (b) 25, and (c) 40GPa. Surface state spectra along $\bar{\mathrm{M}}-\bar{\Gamma}-\bar{\mathrm{M}}$ paths on the (001) plane at (d) 5, (e) 25, and (f) 40GPa. Topological surface states (TSS) are indicated with white arrows.}\label{presstopo}
\end{figure*}

Upon conducting a more detailed analysis of the phonon spectra and phDOS, we can find that the phonon spectra under pressure can be well separated into distinct high and low frequency regions. Specifically, the high-frequency region is mainly contributed by kagome Ti atoms, displaying clear kagome phonon flat band and VHSs. On the other hand, the low-frequency region primarily corresponds to the vibrational modes of Cs and Bi atoms. By separating $\lambda$ into two components, namely $\lambda_{\mathrm{Cs+Bi}}$ and $\lambda_{\mathrm{Ti}}$, we observe that $\lambda_{\mathrm{Ti}}$ gradually decreases with increasing pressure due to the aforementioned background effect, while $\lambda_{\mathrm{Cs+Bi}}$ exhibits exactly the same trend as $\lambda$, as seen in Fig. S4(a). Consequently, our analysis focuses on examining the low-frequency region within the phonon spectra in order to elucidate the trend of $\lambda$.

The total EPC $\lambda$ can be decomposed into the average coupling strengths $\lambda_{\mathbf{q} \nu}$ over the BZ with $\lambda=\sum_{\mathbf{q},\nu}\lambda_{\mathbf{q} \nu}$. By coloring the phonon spectra with $\lambda _{qv} $, we find that the high $\lambda_{\mathbf{q} \nu}$ are mainly concentrated in the lowest acoustic branch. Taking vibration modes of the lowest acoustic branch near $\Gamma$ and M point as examples in Fig.~\ref{presssup}(g), we observe that these modes primarily involve vibrations along the c-axis. Thus, based on our calculations, we can infer that the anomalous increase of $\lambda$ within 10-20 GPa originates from the contributions of $\lambda_{\mathbf{q} \nu}$ associated with the out-of-plane atomic vibrations.

Next, we analyze superconductivity under pressure from a structural perspective. The vertical distance between Bi2 atoms in adjacent Ti$_3$Bi$_5$ sheets is defined as d$ _{\mathrm{Bi2}} $ as illustrated in Fig.~\ref{band}(a). Because of the quasi-2D structural features of CsTi$_3$Bi$_5$, d$ _{\mathrm{Bi2}} $ is quite large, and interlayer interactions are weak. At low pressures, d(d$ _{\mathrm{Bi2}} $)/dP exhibits significant values and changes dramatically. However, as the pressure exceeds 10GPa, the rate of change in d$ _{\mathrm{Bi2}} $ starts to slow down. Furthermore, when P$>$20 GPa, it approaches a nearly constant value. This trend is also reflected in the ratio of the out-of-plane and in-of-plane lattice parameters c/a as depicted in Fig. S4(b). At 20GPa, d$ _{\mathrm{Bi2}} $ becomes approximately equal to the Bi atomic diameter of 3.2\AA, indicating the formation of covalent bonds among the Bi2 atoms around this pressure. These results demonstrate that the structure undergoes a quasi-2D compression below 10GPa. The weak coupling between Cs atoms and Ti$_3$Bi$_5$ sheets leads to a pronounced response of the out-of-plane lattice parameter c to applied pressure. At 10GPa, the Bi atoms of neighboring Ti$_3$Bi$_5$ sheets start to interact, leading to a significant enhancement of $\lambda_{\mathbf{q} \nu}$ associated with out-of-plane vibrations. However, as the pressure exceeds 20GPa, the structure exhibits a 3D isotropic compression characteristic. The interaction between Bi2 atoms no longer increases but gradually decreases due to the underlying background effect mentioned earlier.

\subsection{Topology Surface States under Pressure}

Next, we explore the effect of pressure on electronic structures of CsTi$_3$Bi$_5$. We illustrate the calculation results at pressures of 5, 25 and 40GPa in Figs.~\ref{presstopo}(a-c), which present the electronic band structures of CsTi$_3$Bi$_5$ with SOC. The increased interlayer interaction under pressure profoundly impacts the band structures, leading to notable changes in the energy levels of VHSs and flat band, particularly at higher pressures. With increasing pressure, both VHSs and flat band gradually shift downwards. Remarkably, the VHSs progressively approach the vicinity of the Fermi level under pressure. Exploring the potential nematicity and other phases induced by VHSs through pressure engineering may be of significant interest.

The non-trivial TSS near $\Gamma$ point of CsTi$_3$Bi$_5$ have been experimentally detected \cite{RN967, RN968, RN970}. However, these TSS are approximately 0.8eV below the Fermi level, limiting their substantial effects on transport and other related properties. By examining the surface state spectra of CsTi$_3$Bi$_5$ under pressure in Figs.~\ref{presstopo}(d-f), we observe the emergence of multiple TSS at time-reversal invariant momenta $\bar{\mathrm{M}}$ and $\bar{\Gamma}$, many of which are located in proximity to the Fermi level. Under pressure, these TSS undergo slight energy shifts. Importantly, once these TSS emerge at a specific pressure, further increases in pressure do not eliminate them. The TSS in CsTi$_3$Bi$_5$ exhibits high tunability and sensitivity to pressure, primarily due to the dramatic changes in electronic structures.

\section{Superconductivity of R$\lowercase{\textbf{b}}$T$\lowercase{\textbf{i}}$$ _{3} $B$\lowercase{\textbf{i}}$$ _{5}$ under doping}

RbTi$ _{3} $Bi$ _{3} $ has similar electronic structure and phonon spectrum to CsTi$ _{3} $Bi$ _{3} $ as shown in Fig. S5. Therefore, we can expect RbTi$_3$Bi$_3$ and CsTi$_3$Bi$_3$ to share similar topological, transport, and superconducting properties. Since the $\mathrm{ T_{c}}$ of RbTi$_3$Bi$_5$ is slightly higher than that of CsTi$_3$Bi$_5$ in our calculations, in this section, we focus on a comprehensive investigation of superconductivity in RbTi$_3$Bi$_5 $ under doping.

\begin{figure*}[t]
	\centering
	\includegraphics[scale=0.405,angle=0]{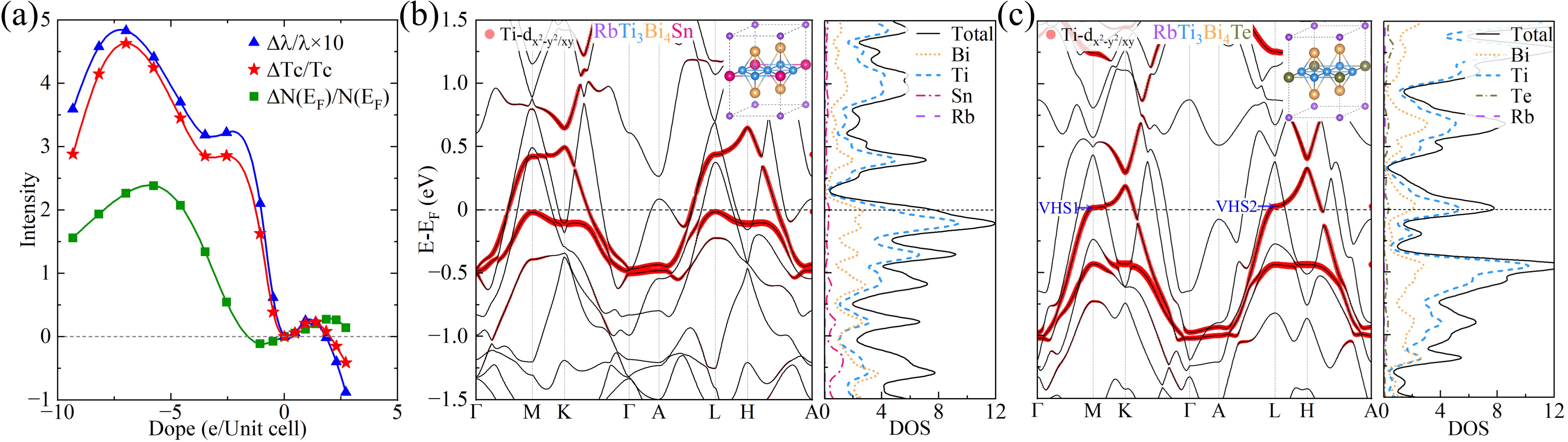}\\
	\caption{\textbf{Superconductivity and electronic structures of RbTi$_\mathbf{3}$Bi$_\mathbf{5}$ under doping.} (a) Relative changes of $\lambda$, $\mathrm{ T_{c}}$, and $\mathrm{N}\left(\mathrm{E}_{\mathrm{F}}\right)$ under doping using the ragid band model: $\Delta \lambda/\lambda$, $\Delta\mathrm{ T_{c} }/\mathrm{ T_{c} }$, and $\Delta \mathrm{N}\left(\mathrm{E}_{\mathrm{F}}\right)/\mathrm{N}\left(\mathrm{E}_{\mathrm{F}}\right)$. Calculated electronic band structures projected by Ti-d$ _{x^{2}-y^{2}/xy} $ orbitals and partial DOS with SOC for (b) RbTi$_3$Bi$_4$Sn and (c) RbTi$_3$Bi$_4$Te, where the insets show the doped structures. Two kagome VHSs near the Fermi level of RbTi$_3$Bi$_4$Te are labeled as “VHS1” and “VHS2”, respectively.}\label{figdope}
\end{figure*}

\begin{table*}\scriptsize
	\renewcommand\arraystretch{1.15}
	\caption{\textbf{Superconductivity of doped sturcutres for RbTi$_\mathbf{3}$Bi$_\mathbf{5}$.} Logarithmic average frequency $\omega$$_l$$_o$$_g$ (K), EPC $\lambda$($\omega$ = $\infty$) and $\mathrm{ T_{c}}$ of doped structures.}
	{\centering
		\begin{tabular}{lp{1.2cm}<{\centering}p{1.5cm}<{\centering}p{1.6cm}<{\centering}p{1.6cm}<{\centering}p{1.6cm}<{\centering}p{1.6cm}<{\centering}p{1.6cm}<{\centering}p{1.6cm}<{\centering}p{1.6cm}<{\centering}p{1.6cm}<{\centering}p{1.2cm}}
			\hline
			\hline
			& RbTi$ _{3} $Bi$ _{5} $ &\multicolumn{1}{l}{Rb\textcolor[rgb]{0,0,1}{V}Ti$ _{2} $Bi$ _{5} $} &\multicolumn{1}{l}{Rb\textcolor[rgb]{0,0,1}{Zr}Ti$ _{2} $Bi$ _{5} $} &\multicolumn{1}{l}{Rb\textcolor[rgb]{0,0,1}{Nb}Ti$ _{2} $Bi$ _{5} $} & \multicolumn{1}{l}{RbTi$ _{3} $Bi$ _{4} $\textcolor[rgb]{0,0,1}{Te}} & \multicolumn{1}{l}{RbTi$ _{3} $Bi$ _{4} $\textcolor[rgb]{0,0,1}{Sb}} & \multicolumn{1}{l}{RbTi$ _{3} $Bi$ _{4} $\textcolor[rgb]{0,0,1}{Sn}} & \multicolumn{1}{l}{RbTi$ _{3} $Bi$ _{4} $\textcolor[rgb]{0,0,1}{Pb}} & \multicolumn{1}{l}{RbTi$ _{3} $Bi$ _{3} $\textcolor[rgb]{0,0,1}{Sn$ _{2} $}} & \multicolumn{1}{l}{RbTi$ _{3} $Bi$ _{3} $\textcolor[rgb]{0,0,1}{Pb$ _{2} $}} \\
			\hline
			$\lambda$ & 0.520 & 0.578 & 0.525 & 0.586 & 0.526 & 0.513 & 0.605 & 0.608 & 0.612 & 0.622 \\
			$\omega_{log}$(K) & 135.8 & 131.3 & 113.7 & 126.5 & 137.7 & 144.9 & 124.6 & 117.8 & 127.6 & 119.1 \\
			$\mathrm{ T_{c}}$(K) & 1.92 & 2.67 & 2.49 & 2.69 & 2.04 & 1.94 & 2.92 & 2.81 & 3.09 & 3.03\\    
			\hline
			\hline	
	\end{tabular}}\label{dope}
\end{table*}

Doping is an effective method to tune electronic structures and superconductivity. Extensive studies have successfully achieved doping in AV$_3$Sb$_5$ to manipulate the competition between superconductivity, CDW, and nematicity \cite{RN1203, RN1158, RN949, RN204, RN1241, RN204, RN1239, RN1014, RN1240, RN1238, RN202}. Notably, the concentration of Sb replacement doping with Sn in AV$_3$Sb$_5$ can reach up to 30\% \cite{RN204}. With reference to AV$_3$Sb$_5$, doping in ATi$_{3}$Bi$_{5}$ is highly feasible. We first employ a rigid band model to analyze the effect of electron and hole doping on superconductivity. Fig.~\ref{figdope}(a) illustrates the relative changes in $\lambda$, $\mathrm{ T_{c}}$, and $\mathrm{N}\left(\mathrm{E}_{\mathrm{F}}\right)$ under doping. Both electron and hole doping result in enhancements of $\mathrm{N}\left(\mathrm{E}_{\mathrm{F}}\right)$ due to the movement of VHSs and flat band towards the Fermi level. As a consequence, there are significant improvements in $\lambda$ and $\mathrm{ T_{c}}$. Especially under hole doping, the highest $\mathrm{N}\left(\mathrm{E}_{\mathrm{F}}\right)$ can be increased by nearly 2.5 times, and $\mathrm{ T_{c}}$ can be increased by nearly 5 times.

In addition to the rigid band model, we also explore specific element substitution doping. Alkali metal substitutions of ATi$_3$Bi$_5$, including LiTi$_3$Bi$_5$, NaTi$_3$Bi$_5$, and KTi$_3$Bi$_5$, are found to be unstable according to our previous study \cite{RN985}. Instead, we substitute one Ti or Bi atom in each unit cell of ATi$_3$Bi$_5$, corresponding to 33\% Ti or 20\% Bi substitution. Ti is replaced by its neighboring elements (Sc, V, Y, Zr and Nb), while Bi is substituted by Ga, Sn, Sb, Te, Pb and Bi. As illustrated in Fig.~\ref{band}(a), the Bi atoms occupy two Wyckoff positions that need to be distinguished. Taking Te doping as an example, we denote the substitution of Bi2 as RbTi$_3$TeBi$_4$, and the substitution of Bi1 as RbTi$_3$Bi$_4$Te as shown in Fig. S6(a). For all doped structures, we first carry out fully geometric relaxation and self-consistent iterations in nonmagnetic (NM), ferromagnetic (FM), and antiferromagnetic (AFM) configurations as seen in Figs. S6(b) and (c). The energies of the three magnetic configurations of all structures are listed in Table S1, which determines the ground state magnetic configuration. Among the doped structures with Bi substitution, the energy of the replacement of Bi2 is consistently higher than that of Bi1. This suggests that the impurity prefers to substitute the Wyckoff position of Bi1. In addition, Sc, Y, and Ga doped CsTi$_3$Bi$_5$ and RbTi$_3$Bi$_5$ all exhibit ferromagnetism, which suppresses superconductivity. Therefore, these structures are not considered in the subsequent analysis.

For these doped structures, we calculate their superconductivity as summarized in Table~\ref{dope}. Several doped structures show varying degrees of improvement compared to the original phase's $\mathrm{ T_{c}}$. $\mathrm{ T_{c}}$ of RbTi$_3$Bi$_4$Sn and RbTi$_3$Bi$_4$Pb under hole doping can be raised to around 3K. However, when increasing the doping concentration to 40$\%$ for Sn and Pb doping, denoted as RbTi$_3$Bi$_3$Sn$ _{2} $ and RbTi$_3$Bi$ _{3} $Pb$ _{2} $, the results indicate that there is no significant increase in $\mathrm{ T_{c}}$'s. The electronic bands and DOS of these doped structures are given in Figs. S7-22. The kagome VHSs and flat band originating from Ti in-plane orbitals in the band structures are prominent, which undergo significant changes upon Ti doping but relatively maintain structural characteristics in the cases of Bi doping. As examples of hole and electron doping, respectively, the electronic energy bands and DOS of RbTi$_3$Bi$_4$Sn and RbTi$_3$Bi$_4$Te are plotted in Figs.~\ref{figdope}(b) and (c). The kagome flat band of RbTi$_3$Bi$_4$Sn and the VHSs of RbTi$_3$Bi$_4$Te are shifted close to the Fermi level, and the DOS originating from the flat band and VHSs exhibit pronounced peaks near the Fermi level. This indicates a shift of the Fermi level while maintaining the character of kagome band structures. These results are consistent with the above rigid band model, demonstrating that both electron and hole doping can improve $\mathrm{ T_{c}}$ by tuning the kagome electronic structures and increasing the DOS through doping. The effective modulation of superconducting properties and kagome electronic structures by doping highlights the need for further experimental investigations in this system.

\section{Discussion}	
The two superconducting domes observed in AV$_3$Sb$_5$ at low pressures are commonly believed to be related to the competition of CDW phases \cite{RN926, RN9, RN54}. However, theoretical analyses of the superconducting dome at around 20-40GPa are limited \cite{RN136, RN190, RN44}. The properties of CsTi$_3$Bi$_5$ and AV$_3$Sb$_5$ at high pressure, in the absence of CDW, are similar, suggesting that our explanations for the superconducting dome and valley in CsTi$_3$Bi$_5$ may also apply to AV$_3$Sb$_5$. Recent research attributes the appearance of high-pressure superconducting dome in AV$_3$Sb$_5$ to transitions from hexagonal to monoclinic structures \cite{RN628}. In contrast, our work demonstrates that the appearance of a similar dome in CsTi$_3$Bi$_5$ does not rely on structural phase transitions but rather originates from the transition from quasi-2D to 3D compression. Further investigations are needed to explore the similarities and differences in the superconducting properties of ATi$_3$Bi$_5$ and AV$_3$Sb$_5$ under pressure.

Fu and Kane proposed that the proximity effect between the helical Dirac TSS of a topological insulator and an s-wave superconductor can induce Majorana zero modes, which can be utilized for topological quantum calculations \cite{RN440}. This effect has been achieved in CsV$ _{3} $Sb$ _{5} $ \cite{RN1}. Based on our calculations, pressure effectively induces abundant Dirac TSS near the Fermi level, whose combination with superconductivity theoretically gives rise to Majorana zero modes. This opens up another fascinating research topic for ATi$ _{3} $Bi$ _{5} $ and its potential for exploring topological quantum phenomena.

The interlayer van der Waals forces presented in ATi$_3$Bi$_5$ facilitate their mechanical exfoliation from the bulk structures. We explore the structural stability and superconducting properties of CsTi$_3$Bi$_5$ in the 2D monolayer limit. Three slab structures, namely CsTi$_3$Bi$_5$ (slab1), Ti$_3$Bi$_5$ (slab2) and Cs$ _{2} $Ti$_3$Bi$_5$ (slab3), with different combinations of Ti$_3$Bi$_5$ layer and Cs layer are considered as seen in Fig. S23. Phonon calculations show that slab2 has minimal imaginary frequencies, while slab1 and slab3 display good dynamic stability. However, superconducting calculations reveal that their $\mathrm{ T_{c}}$'s are nearly equal to that of the bulk structures.

\section{Summary}
	
In summary, we preemptively address superconducting and topological properties under pressure and doping for ATi$_3$Bi$_5$, as well as their thermal conductivity, heat capacity, electrical resistance, and SHC using first-principles calculations. The estimated $\mathrm{ T_{c}}$ of CsTi$_3$Bi$_5$ presents a special down-up-down trend with increasing pressure. The two decreasing sections can be attributed to the decrease in EPC resulting from reduced electronic DOS and $\left\langle\omega^2\right\rangle$ with increasing pressure. Within this overall decreasing trend, the abnormally increased part is associated with the sharply enhanced EPC in the low-frequency region induced by vertical vibration modes, which is related to the transition from anisotropic to isotropic compression from a structural perspective. With rigid band model calculations under doping, $\mathrm{ T_{c}}$ of RbTi$_3$Bi$_5$ can increase by nearly 5 times accompanied by increased electronic DOS and EPC. Calculations for different atomic substitution structures in RbTi$_3$Bi$_5$ show improvements of $\mathrm{ T_{c}}$ from 1.9 to 3.1K. Furthermore, doping and pressure serve as effective means to tune the kagome VHSs and flat band positions close to $\mathrm{E}_{\mathrm{F}}$. Pressure can also induce abundant Dirac-type TSS at $\mathrm{E}_{\mathrm{F}}$. Additionally, our transport calculations agree well with recent experimental results, indicating the absence of CDW. The resistance of CsTi$_3$Bi$_5$ shows normal metal behavior, and its thermal transport is predominantly governed by electronic contributions. The SHC of CsTi$_3$Bi$_5$ can reach up to 226$ \hbar $·(e·$ \Omega $·cm)$ ^{-1} $ near the Fermi level. Our comprehensive study of the diverse physical properties in ATi$_3$Bi$_5$ underscores the need for further experimental exploration in this field.
	
	\section* {Acknowledgments}
This work is supported in part by the National Key R$ \& $D Program of China (Grant No. 2018YFA0305800), the Strategic Priority Research Program of the Chinese Academy of Sciences (Grants No. XDB28000000), the National Natural Science Foundation of China (Grant No.11834014), and the Innovation Program for Quantum Science and Technology (No. 2021ZD0301800). B.G. is supported in part by the National Natural Science Foundation of China (Grant No. c), the Chinese Academy of Sciences (Grants No. YSBR-030 and No. Y929013EA2), the Strategic Priority Research Program of Chinese Academy of Sciences (Grant No. XDB33000000), and the Beijing Natural Science Foundation (Grant No. Z190011).\\
\\

	\section* {Conflict of interest}
The authors declare no competing interests.
	
	\section* {DATA AVAILABILITY}
The data that support the findings of this study are available from the corresponding authors upon reasonable request.
	\section* {Author Contributions}	
J.Y.Y, B. G. and G. S. designed and supervised the research. X.W.Y. performed theoretical calculations. All authors participated in analyzing results. X.W.Y., J.Y.Y., B.G., and G.S. prepared the figures and manuscript.
	

\begin{thebibliography}{71}%
	\makeatletter
	\providecommand \@ifxundefined [1]{%
		\@ifx{#1\undefined}
	}%
	\providecommand \@ifnum [1]{%
		\ifnum #1\expandafter \@firstoftwo
		\else \expandafter \@secondoftwo
		\fi
	}%
	\providecommand \@ifx [1]{%
		\ifx #1\expandafter \@firstoftwo
		\else \expandafter \@secondoftwo
		\fi
	}%
	\providecommand \natexlab [1]{#1}%
	\providecommand \enquote  [1]{``#1''}%
	\providecommand \bibnamefont  [1]{#1}%
	\providecommand \bibfnamefont [1]{#1}%
	\providecommand \citenamefont [1]{#1}%
	\providecommand \href@noop [0]{\@secondoftwo}%
	\providecommand \href [0]{\begingroup \@sanitize@url \@href}%
	\providecommand \@href[1]{\@@startlink{#1}\@@href}%
	\providecommand \@@href[1]{\endgroup#1\@@endlink}%
	\providecommand \@sanitize@url [0]{\catcode `\\12\catcode `\$12\catcode
		`\&12\catcode `\#12\catcode `\^12\catcode `\_12\catcode `\%12\relax}%
	\providecommand \@@startlink[1]{}%
	\providecommand \@@endlink[0]{}%
	\providecommand \url  [0]{\begingroup\@sanitize@url \@url }%
	\providecommand \@url [1]{\endgroup\@href {#1}{\urlprefix }}%
	\providecommand \urlprefix  [0]{URL }%
	\providecommand \Eprint [0]{\href }%
	\providecommand \doibase [0]{https://doi.org/}%
	\providecommand \selectlanguage [0]{\@gobble}%
	\providecommand \bibinfo  [0]{\@secondoftwo}%
	\providecommand \bibfield  [0]{\@secondoftwo}%
	\providecommand \translation [1]{[#1]}%
	\providecommand \BibitemOpen [0]{}%
	\providecommand \bibitemStop [0]{}%
	\providecommand \bibitemNoStop [0]{.\EOS\space}%
	\providecommand \EOS [0]{\spacefactor3000\relax}%
	\providecommand \BibitemShut  [1]{\csname bibitem#1\endcsname}%
	\let\auto@bib@innerbib\@empty
	\bibitem [{\citenamefont {Syozi}(1951)}]{RN1235}%
	\BibitemOpen
	\bibfield  {author} {\bibinfo {author} {\bibfnamefont {I.}~\bibnamefont
			{Syozi}},\ }\bibfield  {title} {\bibinfo {title} {Statistics of kagome
			lattice},\ }\href {https://doi.org/10.1143/ptp/6.3.306} {\bibfield  {journal}
		{\bibinfo  {journal} {Prog. Theor. Phys.}\ }\textbf {\bibinfo {volume} {6}},\
		\bibinfo {pages} {306} (\bibinfo {year} {1951})}\BibitemShut {NoStop}%
	\bibitem [{\citenamefont {Yin}\ \emph {et~al.}(2022)\citenamefont {Yin},
		\citenamefont {Lian},\ and\ \citenamefont {Hasan}}]{RN976}%
	\BibitemOpen
	\bibfield  {author} {\bibinfo {author} {\bibfnamefont {J.-X.}\ \bibnamefont
			{Yin}}, \bibinfo {author} {\bibfnamefont {B.}~\bibnamefont {Lian}},\ and\
		\bibinfo {author} {\bibfnamefont {M.~Z.}\ \bibnamefont {Hasan}},\ }\bibfield
	{title} {\bibinfo {title} {Topological kagome magnets and superconductors},\
	}\href {https://doi.org/10.1038/s41586-022-05516-0} {\bibfield  {journal}
		{\bibinfo  {journal} {Nature}\ }\textbf {\bibinfo {volume} {612}},\ \bibinfo
		{pages} {647} (\bibinfo {year} {2022})}\BibitemShut {NoStop}%
	\bibitem [{\citenamefont {Yin}\ \emph {et~al.}(2019)\citenamefont {Yin},
		\citenamefont {Zhang}, \citenamefont {Chang}, \citenamefont {Wang},
		\citenamefont {Tsirkin}, \citenamefont {Guguchia}, \citenamefont {Lian},
		\citenamefont {Zhou}, \citenamefont {Jiang}, \citenamefont {Belopolski},
		\citenamefont {Shumiya}, \citenamefont {Multer}, \citenamefont {Litskevich},
		\citenamefont {Cochran}, \citenamefont {Lin}, \citenamefont {Wang},
		\citenamefont {Neupert}, \citenamefont {Jia}, \citenamefont {Lei},\ and\
		\citenamefont {Hasan}}]{RN100}%
	\BibitemOpen
	\bibfield  {author} {\bibinfo {author} {\bibfnamefont {J.-X.}\ \bibnamefont
			{Yin}}, \bibinfo {author} {\bibfnamefont {S.~S.}\ \bibnamefont {Zhang}},
		\bibinfo {author} {\bibfnamefont {G.}~\bibnamefont {Chang}}, \bibinfo
		{author} {\bibfnamefont {Q.}~\bibnamefont {Wang}}, \bibinfo {author}
		{\bibfnamefont {S.~S.}\ \bibnamefont {Tsirkin}}, \bibinfo {author}
		{\bibfnamefont {Z.}~\bibnamefont {Guguchia}}, \bibinfo {author}
		{\bibfnamefont {B.}~\bibnamefont {Lian}}, \bibinfo {author} {\bibfnamefont
			{H.}~\bibnamefont {Zhou}}, \bibinfo {author} {\bibfnamefont {K.}~\bibnamefont
			{Jiang}}, \bibinfo {author} {\bibfnamefont {I.}~\bibnamefont {Belopolski}},
		\bibinfo {author} {\bibfnamefont {N.}~\bibnamefont {Shumiya}}, \bibinfo
		{author} {\bibfnamefont {D.}~\bibnamefont {Multer}}, \bibinfo {author}
		{\bibfnamefont {M.}~\bibnamefont {Litskevich}}, \bibinfo {author}
		{\bibfnamefont {T.~A.}\ \bibnamefont {Cochran}}, \bibinfo {author}
		{\bibfnamefont {H.}~\bibnamefont {Lin}}, \bibinfo {author} {\bibfnamefont
			{Z.}~\bibnamefont {Wang}}, \bibinfo {author} {\bibfnamefont {T.}~\bibnamefont
			{Neupert}}, \bibinfo {author} {\bibfnamefont {S.}~\bibnamefont {Jia}},
		\bibinfo {author} {\bibfnamefont {H.}~\bibnamefont {Lei}},\ and\ \bibinfo
		{author} {\bibfnamefont {M.~Z.}\ \bibnamefont {Hasan}},\ }\bibfield  {title}
	{\bibinfo {title} {Negative flat band magnetism in a spin–orbit-coupled
			correlated kagome magnet},\ }\href
	{https://doi.org/10.1038/s41567-019-0426-7} {\bibfield  {journal} {\bibinfo
			{journal} {Nat. Phys.}\ }\textbf {\bibinfo {volume} {15}},\ \bibinfo {pages}
		{443} (\bibinfo {year} {2019})}\BibitemShut {NoStop}%
	\bibitem [{\citenamefont {Gong}\ \emph {et~al.}(2014)\citenamefont {Gong},
		\citenamefont {Zhu},\ and\ \citenamefont {Sheng}}]{RN109}%
	\BibitemOpen
	\bibfield  {author} {\bibinfo {author} {\bibfnamefont {S.~S.}\ \bibnamefont
			{Gong}}, \bibinfo {author} {\bibfnamefont {W.}~\bibnamefont {Zhu}},\ and\
		\bibinfo {author} {\bibfnamefont {D.~N.}\ \bibnamefont {Sheng}},\ }\bibfield
	{title} {\bibinfo {title} {Emergent chiral spin liquid: fractional quantum
			{Hall} effect in a kagome heisenberg model},\ }\href
	{https://doi.org/10.1038/srep06317} {\bibfield  {journal} {\bibinfo
			{journal} {Sci Rep}\ }\textbf {\bibinfo {volume} {4}},\ \bibinfo {pages}
		{6317} (\bibinfo {year} {2014})}\BibitemShut {NoStop}%
	\bibitem [{\citenamefont {Tang}\ \emph {et~al.}(2011)\citenamefont {Tang},
		\citenamefont {Mei},\ and\ \citenamefont {Wen}}]{RN110}%
	\BibitemOpen
	\bibfield  {author} {\bibinfo {author} {\bibfnamefont {E.}~\bibnamefont
			{Tang}}, \bibinfo {author} {\bibfnamefont {J.~W.}\ \bibnamefont {Mei}},\ and\
		\bibinfo {author} {\bibfnamefont {X.~G.}\ \bibnamefont {Wen}},\ }\bibfield
	{title} {\bibinfo {title} {High-temperature fractional quantum {Hall}
			states},\ }\href {https://doi.org/10.1103/PhysRevLett.106.236802} {\bibfield
		{journal} {\bibinfo  {journal} {Phys. Rev. Lett.}\ }\textbf {\bibinfo
			{volume} {106}},\ \bibinfo {pages} {236802} (\bibinfo {year}
		{2011})}\BibitemShut {NoStop}%
	\bibitem [{\citenamefont {Jiang}\ \emph {et~al.}(2017)\citenamefont {Jiang},
		\citenamefont {Devereaux},\ and\ \citenamefont {Kivelson}}]{RN116}%
	\BibitemOpen
	\bibfield  {author} {\bibinfo {author} {\bibfnamefont {H.~C.}\ \bibnamefont
			{Jiang}}, \bibinfo {author} {\bibfnamefont {T.}~\bibnamefont {Devereaux}},\
		and\ \bibinfo {author} {\bibfnamefont {S.~A.}\ \bibnamefont {Kivelson}},\
	}\bibfield  {title} {\bibinfo {title} {Holon {Wigner} crystal in a lightly
			doped kagome quantum spin liquid},\ }\href
	{https://doi.org/10.1103/PhysRevLett.119.067002} {\bibfield  {journal}
		{\bibinfo  {journal} {Phys. Rev. Lett.}\ }\textbf {\bibinfo {volume} {119}},\
		\bibinfo {pages} {067002} (\bibinfo {year} {2017})}\BibitemShut {NoStop}%
	\bibitem [{\citenamefont {Miyahara}\ \emph {et~al.}(2007)\citenamefont
		{Miyahara}, \citenamefont {Kusuta},\ and\ \citenamefont {Furukawa}}]{RN1231}%
	\BibitemOpen
	\bibfield  {author} {\bibinfo {author} {\bibfnamefont {S.}~\bibnamefont
			{Miyahara}}, \bibinfo {author} {\bibfnamefont {S.}~\bibnamefont {Kusuta}},\
		and\ \bibinfo {author} {\bibfnamefont {N.}~\bibnamefont {Furukawa}},\
	}\bibfield  {title} {\bibinfo {title} {{BCS} theory on a flat band lattice},\
	}\href {https://doi.org/10.1016/j.physc.2007.03.393} {\bibfield  {journal}
		{\bibinfo  {journal} {Physica C: Superconductivity}\ }\textbf {\bibinfo
			{volume} {460-462}},\ \bibinfo {pages} {1145} (\bibinfo {year}
		{2007})}\BibitemShut {NoStop}%
	\bibitem [{\citenamefont {Ohgushi}\ \emph {et~al.}(2000)\citenamefont
		{Ohgushi}, \citenamefont {Murakami},\ and\ \citenamefont {Nagaosa}}]{RN1229}%
	\BibitemOpen
	\bibfield  {author} {\bibinfo {author} {\bibfnamefont {K.}~\bibnamefont
			{Ohgushi}}, \bibinfo {author} {\bibfnamefont {S.}~\bibnamefont {Murakami}},\
		and\ \bibinfo {author} {\bibfnamefont {N.}~\bibnamefont {Nagaosa}},\
	}\bibfield  {title} {\bibinfo {title} {Spin anisotropy and quantum {Hall}
			effect in the kagome lattice: Chiral spin state based on a ferromagnet},\
	}\href {https://doi.org/DOI 10.1103/PhysRevB.62.R6065} {\bibfield  {journal}
		{\bibinfo  {journal} {Phys. Rev. B}\ }\textbf {\bibinfo {volume} {62}},\
		\bibinfo {pages} {R6065} (\bibinfo {year} {2000})}\BibitemShut {NoStop}%
	\bibitem [{\citenamefont {Guo}\ and\ \citenamefont {Franz}(2009)}]{RN1230}%
	\BibitemOpen
	\bibfield  {author} {\bibinfo {author} {\bibfnamefont {H.~M.}\ \bibnamefont
			{Guo}}\ and\ \bibinfo {author} {\bibfnamefont {M.}~\bibnamefont {Franz}},\
	}\bibfield  {title} {\bibinfo {title} {Topological insulator on the kagome
			lattice},\ }\href {https://doi.org/10.1103/PhysRevB.80.113102} {\bibfield
		{journal} {\bibinfo  {journal} {Phys. Rev. B}\ }\textbf {\bibinfo {volume}
			{80}},\ \bibinfo {pages} {113102} (\bibinfo {year} {2009})}\BibitemShut
	{NoStop}%
	\bibitem [{\citenamefont {Kiesel}\ \emph {et~al.}(2013)\citenamefont {Kiesel},
		\citenamefont {Platt},\ and\ \citenamefont {Thomale}}]{RN16}%
	\BibitemOpen
	\bibfield  {author} {\bibinfo {author} {\bibfnamefont {M.~L.}\ \bibnamefont
			{Kiesel}}, \bibinfo {author} {\bibfnamefont {C.}~\bibnamefont {Platt}},\ and\
		\bibinfo {author} {\bibfnamefont {R.}~\bibnamefont {Thomale}},\ }\bibfield
	{title} {\bibinfo {title} {Unconventional {Fermi} surface instabilities in
			the kagome {Hubbard} model},\ }\href
	{https://doi.org/10.1103/PhysRevLett.110.126405} {\bibfield  {journal}
		{\bibinfo  {journal} {Phys. Rev. Lett.}\ }\textbf {\bibinfo {volume} {110}},\
		\bibinfo {pages} {126405} (\bibinfo {year} {2013})}\BibitemShut {NoStop}%
	\bibitem [{\citenamefont {Kiesel}\ and\ \citenamefont {Thomale}(2012)}]{RN856}%
	\BibitemOpen
	\bibfield  {author} {\bibinfo {author} {\bibfnamefont {M.~L.}\ \bibnamefont
			{Kiesel}}\ and\ \bibinfo {author} {\bibfnamefont {R.}~\bibnamefont
			{Thomale}},\ }\bibfield  {title} {\bibinfo {title} {Sublattice interference
			in the kagome {Hubbard} model},\ }\href
	{https://doi.org/10.1103/PhysRevB.86.121105} {\bibfield  {journal} {\bibinfo
			{journal} {Phys. Rev. B}\ }\textbf {\bibinfo {volume} {86}},\ \bibinfo
		{pages} {121105} (\bibinfo {year} {2012})}\BibitemShut {NoStop}%
	\bibitem [{\citenamefont {Wang}\ \emph {et~al.}(2013)\citenamefont {Wang},
		\citenamefont {Li}, \citenamefont {Xiang},\ and\ \citenamefont
		{Wang}}]{RN19}%
	\BibitemOpen
	\bibfield  {author} {\bibinfo {author} {\bibfnamefont {W.-S.}\ \bibnamefont
			{Wang}}, \bibinfo {author} {\bibfnamefont {Z.-Z.}\ \bibnamefont {Li}},
		\bibinfo {author} {\bibfnamefont {Y.-Y.}\ \bibnamefont {Xiang}},\ and\
		\bibinfo {author} {\bibfnamefont {Q.-H.}\ \bibnamefont {Wang}},\ }\bibfield
	{title} {\bibinfo {title} {Competing electronic orders on kagome lattices at
			van {Hove} filling},\ }\href {https://doi.org/10.1103/PhysRevB.87.115135}
	{\bibfield  {journal} {\bibinfo  {journal} {Phys. Rev. B}\ }\textbf {\bibinfo
			{volume} {87}},\ \bibinfo {pages} {115135} (\bibinfo {year}
		{2013})}\BibitemShut {NoStop}%
	\bibitem [{\citenamefont {Liu}\ \emph {et~al.}(2019)\citenamefont {Liu},
		\citenamefont {Liang}, \citenamefont {Liu}, \citenamefont {Xu}, \citenamefont
		{Li}, \citenamefont {Chen}, \citenamefont {Pei}, \citenamefont {Shi},
		\citenamefont {Mo}, \citenamefont {Dudin}, \citenamefont {Kim}, \citenamefont
		{Cacho}, \citenamefont {Li}, \citenamefont {Sun}, \citenamefont {Yang},
		\citenamefont {Liu}, \citenamefont {Parkin}, \citenamefont {Felser},\ and\
		\citenamefont {Chen}}]{RN104}%
	\BibitemOpen
	\bibfield  {author} {\bibinfo {author} {\bibfnamefont {D.~F.}\ \bibnamefont
			{Liu}}, \bibinfo {author} {\bibfnamefont {A.~J.}\ \bibnamefont {Liang}},
		\bibinfo {author} {\bibfnamefont {E.~K.}\ \bibnamefont {Liu}}, \bibinfo
		{author} {\bibfnamefont {Q.~N.}\ \bibnamefont {Xu}}, \bibinfo {author}
		{\bibfnamefont {Y.~W.}\ \bibnamefont {Li}}, \bibinfo {author} {\bibfnamefont
			{C.}~\bibnamefont {Chen}}, \bibinfo {author} {\bibfnamefont {D.}~\bibnamefont
			{Pei}}, \bibinfo {author} {\bibfnamefont {W.~J.}\ \bibnamefont {Shi}},
		\bibinfo {author} {\bibfnamefont {S.~K.}\ \bibnamefont {Mo}}, \bibinfo
		{author} {\bibfnamefont {P.}~\bibnamefont {Dudin}}, \bibinfo {author}
		{\bibfnamefont {T.}~\bibnamefont {Kim}}, \bibinfo {author} {\bibfnamefont
			{C.}~\bibnamefont {Cacho}}, \bibinfo {author} {\bibfnamefont
			{G.}~\bibnamefont {Li}}, \bibinfo {author} {\bibfnamefont {Y.}~\bibnamefont
			{Sun}}, \bibinfo {author} {\bibfnamefont {L.~X.}\ \bibnamefont {Yang}},
		\bibinfo {author} {\bibfnamefont {Z.~K.}\ \bibnamefont {Liu}}, \bibinfo
		{author} {\bibfnamefont {S.~S.~P.}\ \bibnamefont {Parkin}}, \bibinfo {author}
		{\bibfnamefont {C.}~\bibnamefont {Felser}},\ and\ \bibinfo {author}
		{\bibfnamefont {Y.~L.}\ \bibnamefont {Chen}},\ }\bibfield  {title} {\bibinfo
		{title} {Magnetic {{Weyl}} semimetal phase in a kagome crystal},\ }\href
	{https://doi.org/10.1126/science.aav2873} {\bibfield  {journal} {\bibinfo
			{journal} {Science}\ }\textbf {\bibinfo {volume} {365}},\ \bibinfo {pages}
		{1282} (\bibinfo {year} {2019})}\BibitemShut {NoStop}%
	\bibitem [{\citenamefont {Morali}\ \emph {et~al.}(2019)\citenamefont {Morali},
		\citenamefont {Batabyal}, \citenamefont {Nag}, \citenamefont {Liu},
		\citenamefont {Xu}, \citenamefont {Sun}, \citenamefont {Yan}, \citenamefont
		{Felser}, \citenamefont {Avraham},\ and\ \citenamefont {Beidenkopf}}]{RN502}%
	\BibitemOpen
	\bibfield  {author} {\bibinfo {author} {\bibfnamefont {N.}~\bibnamefont
			{Morali}}, \bibinfo {author} {\bibfnamefont {R.}~\bibnamefont {Batabyal}},
		\bibinfo {author} {\bibfnamefont {P.~K.}\ \bibnamefont {Nag}}, \bibinfo
		{author} {\bibfnamefont {E.}~\bibnamefont {Liu}}, \bibinfo {author}
		{\bibfnamefont {Q.}~\bibnamefont {Xu}}, \bibinfo {author} {\bibfnamefont
			{Y.}~\bibnamefont {Sun}}, \bibinfo {author} {\bibfnamefont {B.}~\bibnamefont
			{Yan}}, \bibinfo {author} {\bibfnamefont {C.}~\bibnamefont {Felser}},
		\bibinfo {author} {\bibfnamefont {N.}~\bibnamefont {Avraham}},\ and\ \bibinfo
		{author} {\bibfnamefont {H.}~\bibnamefont {Beidenkopf}},\ }\bibfield  {title}
	{\bibinfo {title} {{Fermi}-arc diversity on surface terminations of the
			magnetic {{Weyl}} semimetal {Co$_{3}$Sn$_{2}$S$_{2}$}},\ }\href
	{https://doi.org/10.1126/science.aav2334} {\bibfield  {journal} {\bibinfo
			{journal} {Science}\ }\textbf {\bibinfo {volume} {365}},\ \bibinfo {pages}
		{1286} (\bibinfo {year} {2019})}\BibitemShut {NoStop}%
	\bibitem [{\citenamefont {You}\ and\ \citenamefont {Feng}(2023)}]{RN1253}%
	\BibitemOpen
	\bibfield  {author} {\bibinfo {author} {\bibfnamefont {J.~Y.}\ \bibnamefont
			{You}}\ and\ \bibinfo {author} {\bibfnamefont {Y.~P.}\ \bibnamefont {Feng}},\
	}\bibfield  {title} {\bibinfo {title} {A two-dimensional kagome magnet with
			tunable topological phases},\ }\href
	{https://doi.org/10.1016/j.mtchem.2023.101566} {\bibfield  {journal}
		{\bibinfo  {journal} {Materials Today Chemistry}\ }\textbf {\bibinfo {volume}
			{30}},\ \bibinfo {pages} {101566} (\bibinfo {year} {2023})}\BibitemShut
	{NoStop}%
	\bibitem [{\citenamefont {Liu}\ \emph {et~al.}(2018)\citenamefont {Liu},
		\citenamefont {Sun}, \citenamefont {Kumar}, \citenamefont {Muchler},
		\citenamefont {Sun}, \citenamefont {Jiao}, \citenamefont {Yang},
		\citenamefont {Liu}, \citenamefont {Liang}, \citenamefont {Xu}, \citenamefont
		{Kroder}, \citenamefont {Suss}, \citenamefont {Borrmann}, \citenamefont
		{Shekhar}, \citenamefont {Wang}, \citenamefont {Xi}, \citenamefont {Wang},
		\citenamefont {Schnelle}, \citenamefont {Wirth}, \citenamefont {Chen},
		\citenamefont {Goennenwein},\ and\ \citenamefont {Felser}}]{RN99}%
	\BibitemOpen
	\bibfield  {author} {\bibinfo {author} {\bibfnamefont {E.}~\bibnamefont
			{Liu}}, \bibinfo {author} {\bibfnamefont {Y.}~\bibnamefont {Sun}}, \bibinfo
		{author} {\bibfnamefont {N.}~\bibnamefont {Kumar}}, \bibinfo {author}
		{\bibfnamefont {L.}~\bibnamefont {Muchler}}, \bibinfo {author} {\bibfnamefont
			{A.}~\bibnamefont {Sun}}, \bibinfo {author} {\bibfnamefont {L.}~\bibnamefont
			{Jiao}}, \bibinfo {author} {\bibfnamefont {S.~Y.}\ \bibnamefont {Yang}},
		\bibinfo {author} {\bibfnamefont {D.}~\bibnamefont {Liu}}, \bibinfo {author}
		{\bibfnamefont {A.}~\bibnamefont {Liang}}, \bibinfo {author} {\bibfnamefont
			{Q.}~\bibnamefont {Xu}}, \bibinfo {author} {\bibfnamefont {J.}~\bibnamefont
			{Kroder}}, \bibinfo {author} {\bibfnamefont {V.}~\bibnamefont {Suss}},
		\bibinfo {author} {\bibfnamefont {H.}~\bibnamefont {Borrmann}}, \bibinfo
		{author} {\bibfnamefont {C.}~\bibnamefont {Shekhar}}, \bibinfo {author}
		{\bibfnamefont {Z.}~\bibnamefont {Wang}}, \bibinfo {author} {\bibfnamefont
			{C.}~\bibnamefont {Xi}}, \bibinfo {author} {\bibfnamefont {W.}~\bibnamefont
			{Wang}}, \bibinfo {author} {\bibfnamefont {W.}~\bibnamefont {Schnelle}},
		\bibinfo {author} {\bibfnamefont {S.}~\bibnamefont {Wirth}}, \bibinfo
		{author} {\bibfnamefont {Y.}~\bibnamefont {Chen}}, \bibinfo {author}
		{\bibfnamefont {S.~T.~B.}\ \bibnamefont {Goennenwein}},\ and\ \bibinfo
		{author} {\bibfnamefont {C.}~\bibnamefont {Felser}},\ }\bibfield  {title}
	{\bibinfo {title} {Giant anomalous {Hall} effect in a ferromagnetic
			kagome-lattice semimetal},\ }\href
	{https://doi.org/10.1038/s41567-018-0234-5} {\bibfield  {journal} {\bibinfo
			{journal} {Nat. Phys.}\ }\textbf {\bibinfo {volume} {14}},\ \bibinfo {pages}
		{1125} (\bibinfo {year} {2018})}\BibitemShut {NoStop}%
	\bibitem [{\citenamefont {Nayak}\ \emph {et~al.}(2016)\citenamefont {Nayak},
		\citenamefont {Fischer}, \citenamefont {Sun}, \citenamefont {Yan},
		\citenamefont {Karel}, \citenamefont {Komarek}, \citenamefont {Shekhar},
		\citenamefont {Kumar}, \citenamefont {Schnelle}, \citenamefont {Kubler},
		\citenamefont {Felser},\ and\ \citenamefont {Parkin}}]{RN264}%
	\BibitemOpen
	\bibfield  {author} {\bibinfo {author} {\bibfnamefont {A.~K.}\ \bibnamefont
			{Nayak}}, \bibinfo {author} {\bibfnamefont {J.~E.}\ \bibnamefont {Fischer}},
		\bibinfo {author} {\bibfnamefont {Y.}~\bibnamefont {Sun}}, \bibinfo {author}
		{\bibfnamefont {B.}~\bibnamefont {Yan}}, \bibinfo {author} {\bibfnamefont
			{J.}~\bibnamefont {Karel}}, \bibinfo {author} {\bibfnamefont {A.~C.}\
			\bibnamefont {Komarek}}, \bibinfo {author} {\bibfnamefont {C.}~\bibnamefont
			{Shekhar}}, \bibinfo {author} {\bibfnamefont {N.}~\bibnamefont {Kumar}},
		\bibinfo {author} {\bibfnamefont {W.}~\bibnamefont {Schnelle}}, \bibinfo
		{author} {\bibfnamefont {J.}~\bibnamefont {Kubler}}, \bibinfo {author}
		{\bibfnamefont {C.}~\bibnamefont {Felser}},\ and\ \bibinfo {author}
		{\bibfnamefont {S.~S.}\ \bibnamefont {Parkin}},\ }\bibfield  {title}
	{\bibinfo {title} {Large anomalous {Hall} effect driven by a nonvanishing
			berry curvature in the noncolinear antiferromagnet {Mn$_{3}$Ge}},\ }\href
	{https://doi.org/10.1126/sciadv.1501870} {\bibfield  {journal} {\bibinfo
			{journal} {Sci Adv}\ }\textbf {\bibinfo {volume} {2}},\ \bibinfo {pages}
		{e1501870} (\bibinfo {year} {2016})}\BibitemShut {NoStop}%
	\bibitem [{\citenamefont {Nakatsuji}\ \emph {et~al.}(2015)\citenamefont
		{Nakatsuji}, \citenamefont {Kiyohara},\ and\ \citenamefont {Higo}}]{RN92}%
	\BibitemOpen
	\bibfield  {author} {\bibinfo {author} {\bibfnamefont {S.}~\bibnamefont
			{Nakatsuji}}, \bibinfo {author} {\bibfnamefont {N.}~\bibnamefont
			{Kiyohara}},\ and\ \bibinfo {author} {\bibfnamefont {T.}~\bibnamefont
			{Higo}},\ }\bibfield  {title} {\bibinfo {title} {Large anomalous {Hall}
			effect in a non-collinear antiferromagnet at room temperature},\ }\href
	{https://doi.org/10.1038/nature15723} {\bibfield  {journal} {\bibinfo
			{journal} {Nature}\ }\textbf {\bibinfo {volume} {527}},\ \bibinfo {pages}
		{212} (\bibinfo {year} {2015})}\BibitemShut {NoStop}%
	\bibitem [{\citenamefont {Xu}\ \emph {et~al.}(2015)\citenamefont {Xu},
		\citenamefont {Lian},\ and\ \citenamefont {Zhang}}]{RN108}%
	\BibitemOpen
	\bibfield  {author} {\bibinfo {author} {\bibfnamefont {G.}~\bibnamefont
			{Xu}}, \bibinfo {author} {\bibfnamefont {B.}~\bibnamefont {Lian}},\ and\
		\bibinfo {author} {\bibfnamefont {S.~C.}\ \bibnamefont {Zhang}},\ }\bibfield
	{title} {\bibinfo {title} {Intrinsic quantum anomalous {Hall} effect in the
			kagome lattice {Cs}$_{2}${LiMn}$_{3}${F}$_{12}$},\ }\href
	{https://doi.org/10.1103/PhysRevLett.115.186802} {\bibfield  {journal}
		{\bibinfo  {journal} {Phys. Rev. Lett.}\ }\textbf {\bibinfo {volume} {115}},\
		\bibinfo {pages} {186802} (\bibinfo {year} {2015})}\BibitemShut {NoStop}%
	\bibitem [{\citenamefont {Zhang}\ \emph
		{et~al.}(2021{\natexlab{a}})\citenamefont {Zhang}, \citenamefont {You},
		\citenamefont {Ma}, \citenamefont {Gu},\ and\ \citenamefont {Su}}]{RN232}%
	\BibitemOpen
	\bibfield  {author} {\bibinfo {author} {\bibfnamefont {Z.}~\bibnamefont
			{Zhang}}, \bibinfo {author} {\bibfnamefont {J.-Y.}\ \bibnamefont {You}},
		\bibinfo {author} {\bibfnamefont {X.-Y.}\ \bibnamefont {Ma}}, \bibinfo
		{author} {\bibfnamefont {B.}~\bibnamefont {Gu}},\ and\ \bibinfo {author}
		{\bibfnamefont {G.}~\bibnamefont {Su}},\ }\bibfield  {title} {\bibinfo
		{title} {Kagome quantum anomalous {Hall} effect with high chern number and
			large band gap},\ }\href {https://doi.org/10.1103/PhysRevB.103.014410}
	{\bibfield  {journal} {\bibinfo  {journal} {Phys. Rev. B}\ }\textbf {\bibinfo
			{volume} {103}},\ \bibinfo {pages} {014410} (\bibinfo {year}
		{2021}{\natexlab{a}})}\BibitemShut {NoStop}%
	\bibitem [{\citenamefont {Baidya}\ \emph {et~al.}(2020)\citenamefont {Baidya},
		\citenamefont {Mallik}, \citenamefont {Bhattacharjee},\ and\ \citenamefont
		{Saha-Dasgupta}}]{RN253}%
	\BibitemOpen
	\bibfield  {author} {\bibinfo {author} {\bibfnamefont {S.}~\bibnamefont
			{Baidya}}, \bibinfo {author} {\bibfnamefont {A.~V.}\ \bibnamefont {Mallik}},
		\bibinfo {author} {\bibfnamefont {S.}~\bibnamefont {Bhattacharjee}},\ and\
		\bibinfo {author} {\bibfnamefont {T.}~\bibnamefont {Saha-Dasgupta}},\
	}\bibfield  {title} {\bibinfo {title} {Interplay of magnetism and topological
			superconductivity in bilayer kagome metals},\ }\href
	{https://doi.org/10.1103/PhysRevLett.125.026401} {\bibfield  {journal}
		{\bibinfo  {journal} {Phys. Rev. Lett.}\ }\textbf {\bibinfo {volume} {125}},\
		\bibinfo {pages} {026401} (\bibinfo {year} {2020})}\BibitemShut {NoStop}%
	\bibitem [{\citenamefont {You}\ \emph {et~al.}(2022)\citenamefont {You},
		\citenamefont {Gu}, \citenamefont {Su},\ and\ \citenamefont {Feng}}]{RN514}%
	\BibitemOpen
	\bibfield  {author} {\bibinfo {author} {\bibfnamefont {J.~Y.}\ \bibnamefont
			{You}}, \bibinfo {author} {\bibfnamefont {B.}~\bibnamefont {Gu}}, \bibinfo
		{author} {\bibfnamefont {G.}~\bibnamefont {Su}},\ and\ \bibinfo {author}
		{\bibfnamefont {Y.~P.}\ \bibnamefont {Feng}},\ }\bibfield  {title} {\bibinfo
		{title} {Emergent kagome electrides},\ }\href
	{https://doi.org/10.1021/jacs.2c00177} {\bibfield  {journal} {\bibinfo
			{journal} {J. Am. Chem. Soc.}\ }\textbf {\bibinfo {volume} {144}},\ \bibinfo
		{pages} {5527} (\bibinfo {year} {2022})}\BibitemShut {NoStop}%
	\bibitem [{\citenamefont {Ortiz}\ \emph {et~al.}(2019)\citenamefont {Ortiz},
		\citenamefont {Gomes}, \citenamefont {Morey}, \citenamefont {Winiarski},
		\citenamefont {Bordelon}, \citenamefont {Mangum}, \citenamefont {Oswald},
		\citenamefont {Rodriguez-Rivera}, \citenamefont {Neilson}, \citenamefont
		{Wilson}, \citenamefont {Ertekin}, \citenamefont {McQueen},\ and\
		\citenamefont {Toberer}}]{RN5}%
	\BibitemOpen
	\bibfield  {author} {\bibinfo {author} {\bibfnamefont {B.~R.}\ \bibnamefont
			{Ortiz}}, \bibinfo {author} {\bibfnamefont {L.~C.}\ \bibnamefont {Gomes}},
		\bibinfo {author} {\bibfnamefont {J.~R.}\ \bibnamefont {Morey}}, \bibinfo
		{author} {\bibfnamefont {M.}~\bibnamefont {Winiarski}}, \bibinfo {author}
		{\bibfnamefont {M.}~\bibnamefont {Bordelon}}, \bibinfo {author}
		{\bibfnamefont {J.~S.}\ \bibnamefont {Mangum}}, \bibinfo {author}
		{\bibfnamefont {I.~W.~H.}\ \bibnamefont {Oswald}}, \bibinfo {author}
		{\bibfnamefont {J.~A.}\ \bibnamefont {Rodriguez-Rivera}}, \bibinfo {author}
		{\bibfnamefont {J.~R.}\ \bibnamefont {Neilson}}, \bibinfo {author}
		{\bibfnamefont {S.~D.}\ \bibnamefont {Wilson}}, \bibinfo {author}
		{\bibfnamefont {E.}~\bibnamefont {Ertekin}}, \bibinfo {author} {\bibfnamefont
			{T.~M.}\ \bibnamefont {McQueen}},\ and\ \bibinfo {author} {\bibfnamefont
			{E.~S.}\ \bibnamefont {Toberer}},\ }\bibfield  {title} {\bibinfo {title} {New
			kagome prototype materials: discovery of {KV$_{3}$Sb$_{5}$},
			{RbV$_{3}$Sb$_{5}$}, and {{CsV$_{3}$Sb$_{5}$}}},\ }\href
	{https://doi.org/10.1103/PhysRevMaterials.3.094407} {\bibfield  {journal}
		{\bibinfo  {journal} {Phys. Rev. Mater.}\ }\textbf {\bibinfo {volume} {3}},\
		\bibinfo {pages} {094407} (\bibinfo {year} {2019})}\BibitemShut {NoStop}%
	\bibitem [{\citenamefont {Ortiz}\ \emph {et~al.}(2021)\citenamefont {Ortiz},
		\citenamefont {Sarte}, \citenamefont {Kenney}, \citenamefont {Graf},
		\citenamefont {Teicher}, \citenamefont {Seshadri},\ and\ \citenamefont
		{Wilson}}]{RN11}%
	\BibitemOpen
	\bibfield  {author} {\bibinfo {author} {\bibfnamefont {B.~R.}\ \bibnamefont
			{Ortiz}}, \bibinfo {author} {\bibfnamefont {P.~M.}\ \bibnamefont {Sarte}},
		\bibinfo {author} {\bibfnamefont {E.~M.}\ \bibnamefont {Kenney}}, \bibinfo
		{author} {\bibfnamefont {M.~J.}\ \bibnamefont {Graf}}, \bibinfo {author}
		{\bibfnamefont {S.~M.~L.}\ \bibnamefont {Teicher}}, \bibinfo {author}
		{\bibfnamefont {R.}~\bibnamefont {Seshadri}},\ and\ \bibinfo {author}
		{\bibfnamefont {S.~D.}\ \bibnamefont {Wilson}},\ }\bibfield  {title}
	{\bibinfo {title} {Superconductivity in the {Z$_2$} kagome metal
			{KV$_3$Sb$_5$}},\ }\href {https://doi.org/10.1103/PhysRevMaterials.5.034801}
	{\bibfield  {journal} {\bibinfo  {journal} {Phys. Rev. Mater.}\ }\textbf
		{\bibinfo {volume} {5}},\ \bibinfo {pages} {034801} (\bibinfo {year}
		{2021})}\BibitemShut {NoStop}%
	\bibitem [{\citenamefont {Ortiz}\ \emph {et~al.}(2020)\citenamefont {Ortiz},
		\citenamefont {Teicher}, \citenamefont {Hu}, \citenamefont {Zuo},
		\citenamefont {Sarte}, \citenamefont {Schueller}, \citenamefont {Abeykoon},
		\citenamefont {Krogstad}, \citenamefont {Rosenkranz}, \citenamefont {Osborn},
		\citenamefont {Seshadri}, \citenamefont {Balents}, \citenamefont {He},\ and\
		\citenamefont {Wilson}}]{RN8}%
	\BibitemOpen
	\bibfield  {author} {\bibinfo {author} {\bibfnamefont {B.~R.}\ \bibnamefont
			{Ortiz}}, \bibinfo {author} {\bibfnamefont {S.~M.~L.}\ \bibnamefont
			{Teicher}}, \bibinfo {author} {\bibfnamefont {Y.}~\bibnamefont {Hu}},
		\bibinfo {author} {\bibfnamefont {J.~L.}\ \bibnamefont {Zuo}}, \bibinfo
		{author} {\bibfnamefont {P.~M.}\ \bibnamefont {Sarte}}, \bibinfo {author}
		{\bibfnamefont {E.~C.}\ \bibnamefont {Schueller}}, \bibinfo {author}
		{\bibfnamefont {A.~M.~M.}\ \bibnamefont {Abeykoon}}, \bibinfo {author}
		{\bibfnamefont {M.~J.}\ \bibnamefont {Krogstad}}, \bibinfo {author}
		{\bibfnamefont {S.}~\bibnamefont {Rosenkranz}}, \bibinfo {author}
		{\bibfnamefont {R.}~\bibnamefont {Osborn}}, \bibinfo {author} {\bibfnamefont
			{R.}~\bibnamefont {Seshadri}}, \bibinfo {author} {\bibfnamefont
			{L.}~\bibnamefont {Balents}}, \bibinfo {author} {\bibfnamefont
			{J.}~\bibnamefont {He}},\ and\ \bibinfo {author} {\bibfnamefont {S.~D.}\
			\bibnamefont {Wilson}},\ }\bibfield  {title} {\bibinfo {title}
		{{{CsV$_{3}$Sb$_{5}$}}: A {Z$_{2}$} topological kagome metal with a
			superconducting ground state},\ }\href
	{https://doi.org/10.1103/PhysRevLett.125.247002} {\bibfield  {journal}
		{\bibinfo  {journal} {Phys. Rev. Lett.}\ }\textbf {\bibinfo {volume} {125}},\
		\bibinfo {pages} {247002} (\bibinfo {year} {2020})}\BibitemShut {NoStop}%
	\bibitem [{\citenamefont {Yin}\ \emph {et~al.}(2021)\citenamefont {Yin},
		\citenamefont {Tu}, \citenamefont {Gong}, \citenamefont {Fu}, \citenamefont
		{Yan},\ and\ \citenamefont {Lei}}]{RN64}%
	\BibitemOpen
	\bibfield  {author} {\bibinfo {author} {\bibfnamefont {Q.}~\bibnamefont
			{Yin}}, \bibinfo {author} {\bibfnamefont {Z.}~\bibnamefont {Tu}}, \bibinfo
		{author} {\bibfnamefont {C.}~\bibnamefont {Gong}}, \bibinfo {author}
		{\bibfnamefont {Y.}~\bibnamefont {Fu}}, \bibinfo {author} {\bibfnamefont
			{S.}~\bibnamefont {Yan}},\ and\ \bibinfo {author} {\bibfnamefont
			{H.}~\bibnamefont {Lei}},\ }\bibfield  {title} {\bibinfo {title}
		{Superconductivity and normal-state properties of kagome metal
			{RbV$_3$Sb$_5$} single crystals},\ }\href
	{https://doi.org/10.1088/0256-307x/38/3/037403} {\bibfield  {journal}
		{\bibinfo  {journal} {Chin. Phys. Lett.}\ }\textbf {\bibinfo {volume} {38}},\
		\bibinfo {pages} {037403} (\bibinfo {year} {2021})}\BibitemShut {NoStop}%
	\bibitem [{\citenamefont {Chen}\ \emph
		{et~al.}(2021{\natexlab{a}})\citenamefont {Chen}, \citenamefont {Yang},
		\citenamefont {Hu}, \citenamefont {Zhao}, \citenamefont {Yuan}, \citenamefont
		{Xing}, \citenamefont {Qian}, \citenamefont {Huang}, \citenamefont {Li},
		\citenamefont {Ye}, \citenamefont {Ma}, \citenamefont {Ni}, \citenamefont
		{Zhang}, \citenamefont {Yin}, \citenamefont {Gong}, \citenamefont {Tu},
		\citenamefont {Lei}, \citenamefont {Tan}, \citenamefont {Zhou}, \citenamefont
		{Shen}, \citenamefont {Dong}, \citenamefont {Yan}, \citenamefont {Wang},\
		and\ \citenamefont {Gao}}]{RN34}%
	\BibitemOpen
	\bibfield  {author} {\bibinfo {author} {\bibfnamefont {H.}~\bibnamefont
			{Chen}}, \bibinfo {author} {\bibfnamefont {H.}~\bibnamefont {Yang}}, \bibinfo
		{author} {\bibfnamefont {B.}~\bibnamefont {Hu}}, \bibinfo {author}
		{\bibfnamefont {Z.}~\bibnamefont {Zhao}}, \bibinfo {author} {\bibfnamefont
			{J.}~\bibnamefont {Yuan}}, \bibinfo {author} {\bibfnamefont {Y.}~\bibnamefont
			{Xing}}, \bibinfo {author} {\bibfnamefont {G.}~\bibnamefont {Qian}}, \bibinfo
		{author} {\bibfnamefont {Z.}~\bibnamefont {Huang}}, \bibinfo {author}
		{\bibfnamefont {G.}~\bibnamefont {Li}}, \bibinfo {author} {\bibfnamefont
			{Y.}~\bibnamefont {Ye}}, \bibinfo {author} {\bibfnamefont {S.}~\bibnamefont
			{Ma}}, \bibinfo {author} {\bibfnamefont {S.}~\bibnamefont {Ni}}, \bibinfo
		{author} {\bibfnamefont {H.}~\bibnamefont {Zhang}}, \bibinfo {author}
		{\bibfnamefont {Q.}~\bibnamefont {Yin}}, \bibinfo {author} {\bibfnamefont
			{C.}~\bibnamefont {Gong}}, \bibinfo {author} {\bibfnamefont {Z.}~\bibnamefont
			{Tu}}, \bibinfo {author} {\bibfnamefont {H.}~\bibnamefont {Lei}}, \bibinfo
		{author} {\bibfnamefont {H.}~\bibnamefont {Tan}}, \bibinfo {author}
		{\bibfnamefont {S.}~\bibnamefont {Zhou}}, \bibinfo {author} {\bibfnamefont
			{C.}~\bibnamefont {Shen}}, \bibinfo {author} {\bibfnamefont {X.}~\bibnamefont
			{Dong}}, \bibinfo {author} {\bibfnamefont {B.}~\bibnamefont {Yan}}, \bibinfo
		{author} {\bibfnamefont {Z.}~\bibnamefont {Wang}},\ and\ \bibinfo {author}
		{\bibfnamefont {H.~J.}\ \bibnamefont {Gao}},\ }\bibfield  {title} {\bibinfo
		{title} {Roton pair density wave in a strong-coupling kagome
			superconductor},\ }\href {https://doi.org/10.1038/s41586-021-03983-5}
	{\bibfield  {journal} {\bibinfo  {journal} {Nature}\ }\textbf {\bibinfo
			{volume} {599}},\ \bibinfo {pages} {222} (\bibinfo {year}
		{2021}{\natexlab{a}})}\BibitemShut {NoStop}%
	\bibitem [{\citenamefont {Du}\ \emph {et~al.}(2021)\citenamefont {Du},
		\citenamefont {Luo}, \citenamefont {Ortiz}, \citenamefont {Chen},
		\citenamefont {Duan}, \citenamefont {Zhang}, \citenamefont {Lu},
		\citenamefont {Wilson}, \citenamefont {Song},\ and\ \citenamefont
		{Yuan}}]{RN132}%
	\BibitemOpen
	\bibfield  {author} {\bibinfo {author} {\bibfnamefont {F.}~\bibnamefont
			{Du}}, \bibinfo {author} {\bibfnamefont {S.}~\bibnamefont {Luo}}, \bibinfo
		{author} {\bibfnamefont {B.~R.}\ \bibnamefont {Ortiz}}, \bibinfo {author}
		{\bibfnamefont {Y.}~\bibnamefont {Chen}}, \bibinfo {author} {\bibfnamefont
			{W.}~\bibnamefont {Duan}}, \bibinfo {author} {\bibfnamefont {D.}~\bibnamefont
			{Zhang}}, \bibinfo {author} {\bibfnamefont {X.}~\bibnamefont {Lu}}, \bibinfo
		{author} {\bibfnamefont {S.~D.}\ \bibnamefont {Wilson}}, \bibinfo {author}
		{\bibfnamefont {Y.}~\bibnamefont {Song}},\ and\ \bibinfo {author}
		{\bibfnamefont {H.}~\bibnamefont {Yuan}},\ }\bibfield  {title} {\bibinfo
		{title} {Pressure-induced double superconducting domes and charge instability
			in the kagome metal {KV$_3$Sb$_5$}},\ }\href
	{https://doi.org/10.1103/PhysRevB.103.L220504} {\bibfield  {journal}
		{\bibinfo  {journal} {Phys. Rev. B}\ }\textbf {\bibinfo {volume} {103}},\
		\bibinfo {pages} {L220504} (\bibinfo {year} {2021})}\BibitemShut {NoStop}%
	\bibitem [{\citenamefont {Li}\ \emph {et~al.}(2021)\citenamefont {Li},
		\citenamefont {Zhang}, \citenamefont {Yilmaz}, \citenamefont {Pai},
		\citenamefont {Marvinney}, \citenamefont {Said}, \citenamefont {Yin},
		\citenamefont {Gong}, \citenamefont {Tu}, \citenamefont {Vescovo},
		\citenamefont {Nelson}, \citenamefont {Moore}, \citenamefont {Murakami},
		\citenamefont {Lei}, \citenamefont {Lee}, \citenamefont {Lawrie},\ and\
		\citenamefont {Miao}}]{RN52}%
	\BibitemOpen
	\bibfield  {author} {\bibinfo {author} {\bibfnamefont {H.}~\bibnamefont
			{Li}}, \bibinfo {author} {\bibfnamefont {T.}~\bibnamefont {Zhang}}, \bibinfo
		{author} {\bibfnamefont {T.}~\bibnamefont {Yilmaz}}, \bibinfo {author}
		{\bibfnamefont {Y.}~\bibnamefont {Pai}}, \bibinfo {author} {\bibfnamefont
			{C.}~\bibnamefont {Marvinney}}, \bibinfo {author} {\bibfnamefont
			{A.}~\bibnamefont {Said}}, \bibinfo {author} {\bibfnamefont {Q.}~\bibnamefont
			{Yin}}, \bibinfo {author} {\bibfnamefont {C.}~\bibnamefont {Gong}}, \bibinfo
		{author} {\bibfnamefont {Z.}~\bibnamefont {Tu}}, \bibinfo {author}
		{\bibfnamefont {E.}~\bibnamefont {Vescovo}}, \bibinfo {author} {\bibfnamefont
			{C.}~\bibnamefont {Nelson}}, \bibinfo {author} {\bibfnamefont
			{R.}~\bibnamefont {Moore}}, \bibinfo {author} {\bibfnamefont
			{S.}~\bibnamefont {Murakami}}, \bibinfo {author} {\bibfnamefont
			{H.}~\bibnamefont {Lei}}, \bibinfo {author} {\bibfnamefont {H.}~\bibnamefont
			{Lee}}, \bibinfo {author} {\bibfnamefont {B.}~\bibnamefont {Lawrie}},\ and\
		\bibinfo {author} {\bibfnamefont {H.}~\bibnamefont {Miao}},\ }\bibfield
	{title} {\bibinfo {title} {Observation of unconventional charge density wave
			without acoustic phonon anomaly in kagome superconductors {AV$_3$Sb$_5$}
			{($A$=$Rb$, $Cs$)}},\ }\href {https://doi.org/10.1103/PhysRevX.11.031050}
	{\bibfield  {journal} {\bibinfo  {journal} {Phys. Rev. X}\ }\textbf {\bibinfo
			{volume} {11}},\ \bibinfo {pages} {031050} (\bibinfo {year}
		{2021})}\BibitemShut {NoStop}%
	\bibitem [{\citenamefont {Chen}\ \emph
		{et~al.}(2021{\natexlab{b}})\citenamefont {Chen}, \citenamefont {Wang},
		\citenamefont {Yin}, \citenamefont {Gu}, \citenamefont {Jiang}, \citenamefont
		{Tu}, \citenamefont {Gong}, \citenamefont {Uwatoko}, \citenamefont {Sun},
		\citenamefont {Lei}, \citenamefont {Hu},\ and\ \citenamefont {Cheng}}]{RN9}%
	\BibitemOpen
	\bibfield  {author} {\bibinfo {author} {\bibfnamefont {K.~Y.}\ \bibnamefont
			{Chen}}, \bibinfo {author} {\bibfnamefont {N.~N.}\ \bibnamefont {Wang}},
		\bibinfo {author} {\bibfnamefont {Q.~W.}\ \bibnamefont {Yin}}, \bibinfo
		{author} {\bibfnamefont {Y.~H.}\ \bibnamefont {Gu}}, \bibinfo {author}
		{\bibfnamefont {K.}~\bibnamefont {Jiang}}, \bibinfo {author} {\bibfnamefont
			{Z.~J.}\ \bibnamefont {Tu}}, \bibinfo {author} {\bibfnamefont {C.~S.}\
			\bibnamefont {Gong}}, \bibinfo {author} {\bibfnamefont {Y.}~\bibnamefont
			{Uwatoko}}, \bibinfo {author} {\bibfnamefont {J.~P.}\ \bibnamefont {Sun}},
		\bibinfo {author} {\bibfnamefont {H.~C.}\ \bibnamefont {Lei}}, \bibinfo
		{author} {\bibfnamefont {J.~P.}\ \bibnamefont {Hu}},\ and\ \bibinfo {author}
		{\bibfnamefont {J.~G.}\ \bibnamefont {Cheng}},\ }\bibfield  {title} {\bibinfo
		{title} {Double superconducting dome and triple enhancement of {T$_{c}$} in
			the kagome superconductor {{CsV$_{3}$Sb$_{5}$}} under high pressure},\ }\href
	{https://doi.org/10.1103/PhysRevLett.126.247001} {\bibfield  {journal}
		{\bibinfo  {journal} {Phys. Rev. Lett.}\ }\textbf {\bibinfo {volume} {126}},\
		\bibinfo {pages} {247001} (\bibinfo {year} {2021}{\natexlab{b}})}\BibitemShut
	{NoStop}%
	\bibitem [{\citenamefont {Nie}\ \emph {et~al.}(2022)\citenamefont {Nie},
		\citenamefont {Sun}, \citenamefont {Ma}, \citenamefont {Song}, \citenamefont
		{Zheng}, \citenamefont {Liang}, \citenamefont {Wu}, \citenamefont {Yu},
		\citenamefont {Li}, \citenamefont {Shan}, \citenamefont {Zhao}, \citenamefont
		{Li}, \citenamefont {Kang}, \citenamefont {Wu}, \citenamefont {Zhou},
		\citenamefont {Liu}, \citenamefont {Xiang}, \citenamefont {Ying},
		\citenamefont {Wang}, \citenamefont {Wu},\ and\ \citenamefont
		{Chen}}]{RN187}%
	\BibitemOpen
	\bibfield  {author} {\bibinfo {author} {\bibfnamefont {L.}~\bibnamefont
			{Nie}}, \bibinfo {author} {\bibfnamefont {K.}~\bibnamefont {Sun}}, \bibinfo
		{author} {\bibfnamefont {W.}~\bibnamefont {Ma}}, \bibinfo {author}
		{\bibfnamefont {D.}~\bibnamefont {Song}}, \bibinfo {author} {\bibfnamefont
			{L.}~\bibnamefont {Zheng}}, \bibinfo {author} {\bibfnamefont
			{Z.}~\bibnamefont {Liang}}, \bibinfo {author} {\bibfnamefont
			{P.}~\bibnamefont {Wu}}, \bibinfo {author} {\bibfnamefont {F.}~\bibnamefont
			{Yu}}, \bibinfo {author} {\bibfnamefont {J.}~\bibnamefont {Li}}, \bibinfo
		{author} {\bibfnamefont {M.}~\bibnamefont {Shan}}, \bibinfo {author}
		{\bibfnamefont {D.}~\bibnamefont {Zhao}}, \bibinfo {author} {\bibfnamefont
			{S.}~\bibnamefont {Li}}, \bibinfo {author} {\bibfnamefont {B.}~\bibnamefont
			{Kang}}, \bibinfo {author} {\bibfnamefont {Z.}~\bibnamefont {Wu}}, \bibinfo
		{author} {\bibfnamefont {Y.}~\bibnamefont {Zhou}}, \bibinfo {author}
		{\bibfnamefont {K.}~\bibnamefont {Liu}}, \bibinfo {author} {\bibfnamefont
			{Z.}~\bibnamefont {Xiang}}, \bibinfo {author} {\bibfnamefont
			{J.}~\bibnamefont {Ying}}, \bibinfo {author} {\bibfnamefont {Z.}~\bibnamefont
			{Wang}}, \bibinfo {author} {\bibfnamefont {T.}~\bibnamefont {Wu}},\ and\
		\bibinfo {author} {\bibfnamefont {X.}~\bibnamefont {Chen}},\ }\bibfield
	{title} {\bibinfo {title} {Charge-density-wave-driven electronic nematicity
			in a kagome superconductor},\ }\href
	{https://doi.org/10.1038/s41586-022-04493-8} {\bibfield  {journal} {\bibinfo
			{journal} {Nature}\ }\textbf {\bibinfo {volume} {604}},\ \bibinfo {pages}
		{59} (\bibinfo {year} {2022})}\BibitemShut {NoStop}%
	\bibitem [{\citenamefont {Zheng}\ \emph {et~al.}(2022)\citenamefont {Zheng},
		\citenamefont {Wu}, \citenamefont {Yang}, \citenamefont {Nie}, \citenamefont
		{Shan}, \citenamefont {Sun}, \citenamefont {Song}, \citenamefont {Yu},
		\citenamefont {Li}, \citenamefont {Zhao}, \citenamefont {Li}, \citenamefont
		{Kang}, \citenamefont {Zhou}, \citenamefont {Liu}, \citenamefont {Xiang},
		\citenamefont {Ying}, \citenamefont {Wang}, \citenamefont {Wu},\ and\
		\citenamefont {Chen}}]{RN926}%
	\BibitemOpen
	\bibfield  {author} {\bibinfo {author} {\bibfnamefont {L.}~\bibnamefont
			{Zheng}}, \bibinfo {author} {\bibfnamefont {Z.}~\bibnamefont {Wu}}, \bibinfo
		{author} {\bibfnamefont {Y.}~\bibnamefont {Yang}}, \bibinfo {author}
		{\bibfnamefont {L.}~\bibnamefont {Nie}}, \bibinfo {author} {\bibfnamefont
			{M.}~\bibnamefont {Shan}}, \bibinfo {author} {\bibfnamefont {K.}~\bibnamefont
			{Sun}}, \bibinfo {author} {\bibfnamefont {D.}~\bibnamefont {Song}}, \bibinfo
		{author} {\bibfnamefont {F.}~\bibnamefont {Yu}}, \bibinfo {author}
		{\bibfnamefont {J.}~\bibnamefont {Li}}, \bibinfo {author} {\bibfnamefont
			{D.}~\bibnamefont {Zhao}}, \bibinfo {author} {\bibfnamefont {S.}~\bibnamefont
			{Li}}, \bibinfo {author} {\bibfnamefont {B.}~\bibnamefont {Kang}}, \bibinfo
		{author} {\bibfnamefont {Y.}~\bibnamefont {Zhou}}, \bibinfo {author}
		{\bibfnamefont {K.}~\bibnamefont {Liu}}, \bibinfo {author} {\bibfnamefont
			{Z.}~\bibnamefont {Xiang}}, \bibinfo {author} {\bibfnamefont
			{J.}~\bibnamefont {Ying}}, \bibinfo {author} {\bibfnamefont {Z.}~\bibnamefont
			{Wang}}, \bibinfo {author} {\bibfnamefont {T.}~\bibnamefont {Wu}},\ and\
		\bibinfo {author} {\bibfnamefont {X.}~\bibnamefont {Chen}},\ }\bibfield
	{title} {\bibinfo {title} {Emergent charge order in pressurized kagome
			superconductor {CsV$_{3}$Sb$_{5}$}},\ }\href
	{https://doi.org/10.1038/s41586-022-05351-3} {\bibfield  {journal} {\bibinfo
			{journal} {Nature}\ }\textbf {\bibinfo {volume} {611}},\ \bibinfo {pages}
		{682} (\bibinfo {year} {2022})}\BibitemShut {NoStop}%
	\bibitem [{\citenamefont {Zhao}\ \emph {et~al.}(2021)\citenamefont {Zhao},
		\citenamefont {Li}, \citenamefont {Ortiz}, \citenamefont {Teicher},
		\citenamefont {Park}, \citenamefont {Ye}, \citenamefont {Wang}, \citenamefont
		{Balents}, \citenamefont {Wilson},\ and\ \citenamefont {Zeljkovic}}]{RN222}%
	\BibitemOpen
	\bibfield  {author} {\bibinfo {author} {\bibfnamefont {H.}~\bibnamefont
			{Zhao}}, \bibinfo {author} {\bibfnamefont {H.}~\bibnamefont {Li}}, \bibinfo
		{author} {\bibfnamefont {B.~R.}\ \bibnamefont {Ortiz}}, \bibinfo {author}
		{\bibfnamefont {S.~M.~L.}\ \bibnamefont {Teicher}}, \bibinfo {author}
		{\bibfnamefont {T.}~\bibnamefont {Park}}, \bibinfo {author} {\bibfnamefont
			{M.}~\bibnamefont {Ye}}, \bibinfo {author} {\bibfnamefont {Z.}~\bibnamefont
			{Wang}}, \bibinfo {author} {\bibfnamefont {L.}~\bibnamefont {Balents}},
		\bibinfo {author} {\bibfnamefont {S.~D.}\ \bibnamefont {Wilson}},\ and\
		\bibinfo {author} {\bibfnamefont {I.}~\bibnamefont {Zeljkovic}},\ }\bibfield
	{title} {\bibinfo {title} {Cascade of correlated electron states in the
			kagome superconductor {{CsV$_{3}$Sb$_{5}$}}},\ }\href
	{https://doi.org/10.1038/s41586-021-03946-w} {\bibfield  {journal} {\bibinfo
			{journal} {Nature}\ }\textbf {\bibinfo {volume} {599}},\ \bibinfo {pages}
		{216} (\bibinfo {year} {2021})}\BibitemShut {NoStop}%
	\bibitem [{\citenamefont {Yin}(2023)}]{RN1216}%
	\BibitemOpen
	\bibfield  {author} {\bibinfo {author} {\bibfnamefont {J.~X.}\ \bibnamefont
			{Yin}},\ }\bibfield  {title} {\bibinfo {title} {Quantum dephasing of kagome
			superconductivity},\ }\href {https://doi.org/10.1016/j.scib.2023.02.035}
	{\bibfield  {journal} {\bibinfo  {journal} {Sci. Bull.}\ }\textbf {\bibinfo
			{volume} {68}},\ \bibinfo {pages} {568} (\bibinfo {year} {2023})}\BibitemShut
	{NoStop}%
	\bibitem [{\citenamefont {Zhu}\ \emph {et~al.}(2022)\citenamefont {Zhu},
		\citenamefont {Yang}, \citenamefont {Xia}, \citenamefont {Yin}, \citenamefont
		{Wang}, \citenamefont {Zhao}, \citenamefont {Dai}, \citenamefont {Tu},
		\citenamefont {Song}, \citenamefont {Tao}, \citenamefont {Tu}, \citenamefont
		{Gong}, \citenamefont {Lei}, \citenamefont {Guo},\ and\ \citenamefont
		{Li}}]{RN469}%
	\BibitemOpen
	\bibfield  {author} {\bibinfo {author} {\bibfnamefont {C.~C.}\ \bibnamefont
			{Zhu}}, \bibinfo {author} {\bibfnamefont {X.~F.}\ \bibnamefont {Yang}},
		\bibinfo {author} {\bibfnamefont {W.}~\bibnamefont {Xia}}, \bibinfo {author}
		{\bibfnamefont {Q.~W.}\ \bibnamefont {Yin}}, \bibinfo {author} {\bibfnamefont
			{L.~S.}\ \bibnamefont {Wang}}, \bibinfo {author} {\bibfnamefont {C.~C.}\
			\bibnamefont {Zhao}}, \bibinfo {author} {\bibfnamefont {D.~Z.}\ \bibnamefont
			{Dai}}, \bibinfo {author} {\bibfnamefont {C.~P.}\ \bibnamefont {Tu}},
		\bibinfo {author} {\bibfnamefont {B.~Q.}\ \bibnamefont {Song}}, \bibinfo
		{author} {\bibfnamefont {Z.~C.}\ \bibnamefont {Tao}}, \bibinfo {author}
		{\bibfnamefont {Z.~J.}\ \bibnamefont {Tu}}, \bibinfo {author} {\bibfnamefont
			{C.~S.}\ \bibnamefont {Gong}}, \bibinfo {author} {\bibfnamefont {H.~C.}\
			\bibnamefont {Lei}}, \bibinfo {author} {\bibfnamefont {Y.~F.}\ \bibnamefont
			{Guo}},\ and\ \bibinfo {author} {\bibfnamefont {S.~Y.}\ \bibnamefont {Li}},\
	}\bibfield  {title} {\bibinfo {title} {Double-dome superconductivity under
			pressure in the {V-based} kagome metals {AV$_3$Sb$_5$} {(A=Rb and K)}},\
	}\href {https://doi.org/10.1103/PhysRevB.105.094507} {\bibfield  {journal}
		{\bibinfo  {journal} {Phys. Rev. B}\ }\textbf {\bibinfo {volume} {105}},\
		\bibinfo {pages} {094507} (\bibinfo {year} {2022})}\BibitemShut {NoStop}%
	\bibitem [{\citenamefont {Yu}\ \emph {et~al.}(2021)\citenamefont {Yu},
		\citenamefont {Ma}, \citenamefont {Zhuo}, \citenamefont {Liu}, \citenamefont
		{Wen}, \citenamefont {Lei}, \citenamefont {Ying},\ and\ \citenamefont
		{Chen}}]{RN54}%
	\BibitemOpen
	\bibfield  {author} {\bibinfo {author} {\bibfnamefont {F.~H.}\ \bibnamefont
			{Yu}}, \bibinfo {author} {\bibfnamefont {D.~H.}\ \bibnamefont {Ma}}, \bibinfo
		{author} {\bibfnamefont {W.~Z.}\ \bibnamefont {Zhuo}}, \bibinfo {author}
		{\bibfnamefont {S.~Q.}\ \bibnamefont {Liu}}, \bibinfo {author} {\bibfnamefont
			{X.~K.}\ \bibnamefont {Wen}}, \bibinfo {author} {\bibfnamefont
			{B.}~\bibnamefont {Lei}}, \bibinfo {author} {\bibfnamefont {J.~J.}\
			\bibnamefont {Ying}},\ and\ \bibinfo {author} {\bibfnamefont {X.~H.}\
			\bibnamefont {Chen}},\ }\bibfield  {title} {\bibinfo {title} {Unusual
			competition of superconductivity and charge-density-wave state in a
			compressed topological kagome metal},\ }\href
	{https://doi.org/10.1038/s41467-021-23928-w} {\bibfield  {journal} {\bibinfo
			{journal} {Nat. Commun.}\ }\textbf {\bibinfo {volume} {12}},\ \bibinfo
		{pages} {3645} (\bibinfo {year} {2021})}\BibitemShut {NoStop}%
	\bibitem [{\citenamefont {Zhong}\ \emph {et~al.}(2023)\citenamefont {Zhong},
		\citenamefont {Liu}, \citenamefont {Wu}, \citenamefont {Guguchia},
		\citenamefont {Yin}, \citenamefont {Mine}, \citenamefont {Li}, \citenamefont
		{Najafzadeh}, \citenamefont {Das}, \citenamefont {Mielke}, \citenamefont
		{Khasanov}, \citenamefont {Luetkens}, \citenamefont {Suzuki}, \citenamefont
		{Liu}, \citenamefont {Han}, \citenamefont {Kondo}, \citenamefont {Hu},
		\citenamefont {Shin}, \citenamefont {Wang}, \citenamefont {Shi},
		\citenamefont {Yao},\ and\ \citenamefont {Okazaki}}]{RN1203}%
	\BibitemOpen
	\bibfield  {author} {\bibinfo {author} {\bibfnamefont {Y.}~\bibnamefont
			{Zhong}}, \bibinfo {author} {\bibfnamefont {J.}~\bibnamefont {Liu}}, \bibinfo
		{author} {\bibfnamefont {X.}~\bibnamefont {Wu}}, \bibinfo {author}
		{\bibfnamefont {Z.}~\bibnamefont {Guguchia}}, \bibinfo {author}
		{\bibfnamefont {J.~X.}\ \bibnamefont {Yin}}, \bibinfo {author} {\bibfnamefont
			{A.}~\bibnamefont {Mine}}, \bibinfo {author} {\bibfnamefont {Y.}~\bibnamefont
			{Li}}, \bibinfo {author} {\bibfnamefont {S.}~\bibnamefont {Najafzadeh}},
		\bibinfo {author} {\bibfnamefont {D.}~\bibnamefont {Das}}, \bibinfo {author}
		{\bibfnamefont {r.}~\bibnamefont {Mielke}, \bibfnamefont {C.}}, \bibinfo
		{author} {\bibfnamefont {R.}~\bibnamefont {Khasanov}}, \bibinfo {author}
		{\bibfnamefont {H.}~\bibnamefont {Luetkens}}, \bibinfo {author}
		{\bibfnamefont {T.}~\bibnamefont {Suzuki}}, \bibinfo {author} {\bibfnamefont
			{K.}~\bibnamefont {Liu}}, \bibinfo {author} {\bibfnamefont {X.}~\bibnamefont
			{Han}}, \bibinfo {author} {\bibfnamefont {T.}~\bibnamefont {Kondo}}, \bibinfo
		{author} {\bibfnamefont {J.}~\bibnamefont {Hu}}, \bibinfo {author}
		{\bibfnamefont {S.}~\bibnamefont {Shin}}, \bibinfo {author} {\bibfnamefont
			{Z.}~\bibnamefont {Wang}}, \bibinfo {author} {\bibfnamefont {X.}~\bibnamefont
			{Shi}}, \bibinfo {author} {\bibfnamefont {Y.}~\bibnamefont {Yao}},\ and\
		\bibinfo {author} {\bibfnamefont {K.}~\bibnamefont {Okazaki}},\ }\bibfield
	{title} {\bibinfo {title} {Nodeless electron pairing in
			{CsV$_{3}$Sb$_{5}$}-derived kagome superconductors},\ }\href
	{https://doi.org/10.1038/s41586-023-05907-x} {\bibfield  {journal} {\bibinfo
			{journal} {Nature}\ }\textbf {\bibinfo {volume} {617}},\ \bibinfo {pages}
		{488} (\bibinfo {year} {2023})}\BibitemShut {NoStop}%
	\bibitem [{\citenamefont {Wu}\ \emph {et~al.}(2023)\citenamefont {Wu},
		\citenamefont {Tu}, \citenamefont {Wang}, \citenamefont {Yu}, \citenamefont
		{Li}, \citenamefont {Ma}, \citenamefont {Liang}, \citenamefont {Zhang},
		\citenamefont {Zhang}, \citenamefont {Li}, \citenamefont {Yang},
		\citenamefont {Qiao}, \citenamefont {Ying}, \citenamefont {Wu}, \citenamefont
		{Shan}, \citenamefont {Xiang}, \citenamefont {Wang},\ and\ \citenamefont
		{Chen}}]{RN1158}%
	\BibitemOpen
	\bibfield  {author} {\bibinfo {author} {\bibfnamefont {P.}~\bibnamefont
			{Wu}}, \bibinfo {author} {\bibfnamefont {Y.~B.}\ \bibnamefont {Tu}}, \bibinfo
		{author} {\bibfnamefont {Z.~Y.}\ \bibnamefont {Wang}}, \bibinfo {author}
		{\bibfnamefont {S.~K.}\ \bibnamefont {Yu}}, \bibinfo {author} {\bibfnamefont
			{H.~Y.}\ \bibnamefont {Li}}, \bibinfo {author} {\bibfnamefont {W.~R.}\
			\bibnamefont {Ma}}, \bibinfo {author} {\bibfnamefont {Z.~W.}\ \bibnamefont
			{Liang}}, \bibinfo {author} {\bibfnamefont {Y.~M.}\ \bibnamefont {Zhang}},
		\bibinfo {author} {\bibfnamefont {X.~C.}\ \bibnamefont {Zhang}}, \bibinfo
		{author} {\bibfnamefont {Z.~Y.}\ \bibnamefont {Li}}, \bibinfo {author}
		{\bibfnamefont {Y.}~\bibnamefont {Yang}}, \bibinfo {author} {\bibfnamefont
			{Z.~H.}\ \bibnamefont {Qiao}}, \bibinfo {author} {\bibfnamefont {J.~J.}\
			\bibnamefont {Ying}}, \bibinfo {author} {\bibfnamefont {T.}~\bibnamefont
			{Wu}}, \bibinfo {author} {\bibfnamefont {L.}~\bibnamefont {Shan}}, \bibinfo
		{author} {\bibfnamefont {Z.~J.}\ \bibnamefont {Xiang}}, \bibinfo {author}
		{\bibfnamefont {Z.~Y.}\ \bibnamefont {Wang}},\ and\ \bibinfo {author}
		{\bibfnamefont {X.~H.}\ \bibnamefont {Chen}},\ }\bibfield  {title} {\bibinfo
		{title} {Unidirectional electron-phonon coupling in the nematic state of a
			kagome superconductor},\ }\bibfield  {journal} {\bibinfo  {journal} {Nat.
			Phys.}\ }\href {https://doi.org/10.1038/s41567-023-02031-5}
	{10.1038/s41567-023-02031-5} (\bibinfo {year} {2023})\BibitemShut {NoStop}%
	\bibitem [{\citenamefont {Yang}\ \emph
		{et~al.}(2022{\natexlab{a}})\citenamefont {Yang}, \citenamefont {Huang},
		\citenamefont {Zhang}, \citenamefont {Zhao}, \citenamefont {Shi},
		\citenamefont {Luo}, \citenamefont {Zhao}, \citenamefont {Qian},
		\citenamefont {Tan}, \citenamefont {Hu}, \citenamefont {Zhu}, \citenamefont
		{Lu}, \citenamefont {Zhang}, \citenamefont {Sun}, \citenamefont {Cheng},
		\citenamefont {Shen}, \citenamefont {Lin}, \citenamefont {Yan}, \citenamefont
		{Zhou}, \citenamefont {Wang}, \citenamefont {Pennycook}, \citenamefont
		{Chen}, \citenamefont {Dong}, \citenamefont {Zhou},\ and\ \citenamefont
		{Gao}}]{RN949}%
	\BibitemOpen
	\bibfield  {author} {\bibinfo {author} {\bibfnamefont {H.}~\bibnamefont
			{Yang}}, \bibinfo {author} {\bibfnamefont {Z.}~\bibnamefont {Huang}},
		\bibinfo {author} {\bibfnamefont {Y.}~\bibnamefont {Zhang}}, \bibinfo
		{author} {\bibfnamefont {Z.}~\bibnamefont {Zhao}}, \bibinfo {author}
		{\bibfnamefont {J.}~\bibnamefont {Shi}}, \bibinfo {author} {\bibfnamefont
			{H.}~\bibnamefont {Luo}}, \bibinfo {author} {\bibfnamefont {L.}~\bibnamefont
			{Zhao}}, \bibinfo {author} {\bibfnamefont {G.}~\bibnamefont {Qian}}, \bibinfo
		{author} {\bibfnamefont {H.}~\bibnamefont {Tan}}, \bibinfo {author}
		{\bibfnamefont {B.}~\bibnamefont {Hu}}, \bibinfo {author} {\bibfnamefont
			{K.}~\bibnamefont {Zhu}}, \bibinfo {author} {\bibfnamefont {Z.}~\bibnamefont
			{Lu}}, \bibinfo {author} {\bibfnamefont {H.}~\bibnamefont {Zhang}}, \bibinfo
		{author} {\bibfnamefont {J.}~\bibnamefont {Sun}}, \bibinfo {author}
		{\bibfnamefont {J.}~\bibnamefont {Cheng}}, \bibinfo {author} {\bibfnamefont
			{C.}~\bibnamefont {Shen}}, \bibinfo {author} {\bibfnamefont {X.}~\bibnamefont
			{Lin}}, \bibinfo {author} {\bibfnamefont {B.}~\bibnamefont {Yan}}, \bibinfo
		{author} {\bibfnamefont {X.}~\bibnamefont {Zhou}}, \bibinfo {author}
		{\bibfnamefont {Z.}~\bibnamefont {Wang}}, \bibinfo {author} {\bibfnamefont
			{S.~J.}\ \bibnamefont {Pennycook}}, \bibinfo {author} {\bibfnamefont
			{H.}~\bibnamefont {Chen}}, \bibinfo {author} {\bibfnamefont {X.}~\bibnamefont
			{Dong}}, \bibinfo {author} {\bibfnamefont {W.}~\bibnamefont {Zhou}},\ and\
		\bibinfo {author} {\bibfnamefont {H.-J.}\ \bibnamefont {Gao}},\ }\bibfield
	{title} {\bibinfo {title} {Titanium doped kagome superconductor
			{CsV$_{3-x}$Ti$_{x}$Sb$_{5}$} and two distinct phases},\ }\href
	{https://doi.org/10.1016/j.scib.2022.10.015} {\bibfield  {journal} {\bibinfo
			{journal} {Sci. Bull.}\ }\textbf {\bibinfo {volume} {67}},\ \bibinfo {pages}
		{2176} (\bibinfo {year} {2022}{\natexlab{a}})}\BibitemShut {NoStop}%
	\bibitem [{\citenamefont {Oey}\ \emph {et~al.}(2022{\natexlab{a}})\citenamefont
		{Oey}, \citenamefont {Ortiz}, \citenamefont {Kaboudvand}, \citenamefont
		{Frassineti}, \citenamefont {Garcia}, \citenamefont {Cong}, \citenamefont
		{Sanna}, \citenamefont {Mitrovic}, \citenamefont {Seshadri},\ and\
		\citenamefont {Wilson}}]{RN204}%
	\BibitemOpen
	\bibfield  {author} {\bibinfo {author} {\bibfnamefont {Y.~M.}\ \bibnamefont
			{Oey}}, \bibinfo {author} {\bibfnamefont {B.~R.}\ \bibnamefont {Ortiz}},
		\bibinfo {author} {\bibfnamefont {F.}~\bibnamefont {Kaboudvand}}, \bibinfo
		{author} {\bibfnamefont {J.}~\bibnamefont {Frassineti}}, \bibinfo {author}
		{\bibfnamefont {E.}~\bibnamefont {Garcia}}, \bibinfo {author} {\bibfnamefont
			{R.}~\bibnamefont {Cong}}, \bibinfo {author} {\bibfnamefont {S.}~\bibnamefont
			{Sanna}}, \bibinfo {author} {\bibfnamefont {V.~F.}\ \bibnamefont {Mitrovic}},
		\bibinfo {author} {\bibfnamefont {R.}~\bibnamefont {Seshadri}},\ and\
		\bibinfo {author} {\bibfnamefont {S.~D.}\ \bibnamefont {Wilson}},\ }\bibfield
	{title} {\bibinfo {title} {{Fermi} level tuning and double-dome
			superconductivity in the kagome metal {CsV$_{3}$Sb$_{5-x}$Sn$_{x}$}},\ }\href
	{https://doi.org/10.1103/PhysRevMaterials.6.L041801} {\bibfield  {journal}
		{\bibinfo  {journal} {Phys. Rev. Mater.}\ }\textbf {\bibinfo {volume} {6}},\
		\bibinfo {pages} {L041801} (\bibinfo {year}
		{2022}{\natexlab{a}})}\BibitemShut {NoStop}%
	\bibitem [{\citenamefont {Oey}\ \emph {et~al.}(2022{\natexlab{b}})\citenamefont
		{Oey}, \citenamefont {Kaboudvand}, \citenamefont {Ortiz}, \citenamefont
		{Seshadri},\ and\ \citenamefont {Wilson}}]{RN1241}%
	\BibitemOpen
	\bibfield  {author} {\bibinfo {author} {\bibfnamefont {Y.~M.}\ \bibnamefont
			{Oey}}, \bibinfo {author} {\bibfnamefont {F.}~\bibnamefont {Kaboudvand}},
		\bibinfo {author} {\bibfnamefont {B.~R.}\ \bibnamefont {Ortiz}}, \bibinfo
		{author} {\bibfnamefont {R.}~\bibnamefont {Seshadri}},\ and\ \bibinfo
		{author} {\bibfnamefont {S.~D.}\ \bibnamefont {Wilson}},\ }\bibfield  {title}
	{\bibinfo {title} {Tuning charge density wave order and superconductivity in
			the kagome metals {KV$_{3}$Sb$_{5-x}$Sn$_x$} and
			{RbV$_{3}$Sb$_{5-x}$Sn$_x$}},\ }\href
	{https://doi.org/10.1103/PhysRevMaterials.6.074802} {\bibfield  {journal}
		{\bibinfo  {journal} {Phys. Rev. Mater.}\ }\textbf {\bibinfo {volume} {6}},\
		\bibinfo {pages} {074802} (\bibinfo {year} {2022}{\natexlab{b}})}\BibitemShut
	{NoStop}%
	\bibitem [{\citenamefont {Liu}\ \emph {et~al.}(2022{\natexlab{a}})\citenamefont
		{Liu}, \citenamefont {Han}, \citenamefont {Hu}, \citenamefont {Tu},
		\citenamefont {Zhang}, \citenamefont {Long}, \citenamefont {Hou},
		\citenamefont {Mu},\ and\ \citenamefont {Shan}}]{RN1239}%
	\BibitemOpen
	\bibfield  {author} {\bibinfo {author} {\bibfnamefont {M.}~\bibnamefont
			{Liu}}, \bibinfo {author} {\bibfnamefont {T.}~\bibnamefont {Han}}, \bibinfo
		{author} {\bibfnamefont {X.}~\bibnamefont {Hu}}, \bibinfo {author}
		{\bibfnamefont {Y.}~\bibnamefont {Tu}}, \bibinfo {author} {\bibfnamefont
			{Z.}~\bibnamefont {Zhang}}, \bibinfo {author} {\bibfnamefont
			{M.}~\bibnamefont {Long}}, \bibinfo {author} {\bibfnamefont {X.}~\bibnamefont
			{Hou}}, \bibinfo {author} {\bibfnamefont {Q.}~\bibnamefont {Mu}},\ and\
		\bibinfo {author} {\bibfnamefont {L.}~\bibnamefont {Shan}},\ }\bibfield
	{title} {\bibinfo {title} {Evolution of superconductivity and charge density
			wave through {Ta} and {Mo} doping in {{CsV$_{3}$Sb$_{5}$}}},\ }\href
	{https://doi.org/10.1103/PhysRevB.106.L140501} {\bibfield  {journal}
		{\bibinfo  {journal} {Phys. Rev. B}\ }\textbf {\bibinfo {volume} {106}},\
		\bibinfo {pages} {L140501} (\bibinfo {year}
		{2022}{\natexlab{a}})}\BibitemShut {NoStop}%
	\bibitem [{\citenamefont {Liu}\ \emph {et~al.}(2022{\natexlab{b}})\citenamefont
		{Liu}, \citenamefont {Liu}, \citenamefont {Zhu}, \citenamefont {Ji},
		\citenamefont {Wu}, \citenamefont {Sun}, \citenamefont {Bao}, \citenamefont
		{Jiao}, \citenamefont {Xu}, \citenamefont {Ren},\ and\ \citenamefont
		{Cao}}]{RN1014}%
	\BibitemOpen
	\bibfield  {author} {\bibinfo {author} {\bibfnamefont {Y.}~\bibnamefont
			{Liu}}, \bibinfo {author} {\bibfnamefont {C.-C.}\ \bibnamefont {Liu}},
		\bibinfo {author} {\bibfnamefont {Q.-Q.}\ \bibnamefont {Zhu}}, \bibinfo
		{author} {\bibfnamefont {L.-W.}\ \bibnamefont {Ji}}, \bibinfo {author}
		{\bibfnamefont {S.-Q.}\ \bibnamefont {Wu}}, \bibinfo {author} {\bibfnamefont
			{Y.-L.}\ \bibnamefont {Sun}}, \bibinfo {author} {\bibfnamefont {J.-K.}\
			\bibnamefont {Bao}}, \bibinfo {author} {\bibfnamefont {W.-H.}\ \bibnamefont
			{Jiao}}, \bibinfo {author} {\bibfnamefont {X.-F.}\ \bibnamefont {Xu}},
		\bibinfo {author} {\bibfnamefont {Z.}~\bibnamefont {Ren}},\ and\ \bibinfo
		{author} {\bibfnamefont {G.-H.}\ \bibnamefont {Cao}},\ }\bibfield  {title}
	{\bibinfo {title} {Enhancement of superconductivity and suppression of
			charge-density wave in {As}-doped {{CsV$_{3}$Sb$_{5}$}}},\ }\href
	{https://doi.org/10.1103/PhysRevMaterials.6.124803} {\bibfield  {journal}
		{\bibinfo  {journal} {Phys. Rev. Mater.}\ }\textbf {\bibinfo {volume} {6}},\
		\bibinfo {pages} {124803} (\bibinfo {year} {2022}{\natexlab{b}})}\BibitemShut
	{NoStop}%
	\bibitem [{\citenamefont {Zhou}\ \emph
		{et~al.}(2023{\natexlab{a}})\citenamefont {Zhou}, \citenamefont {Li},
		\citenamefont {Liu}, \citenamefont {Hao}, \citenamefont {Dai}, \citenamefont
		{Wang}, \citenamefont {Yao},\ and\ \citenamefont {Wen}}]{RN1240}%
	\BibitemOpen
	\bibfield  {author} {\bibinfo {author} {\bibfnamefont {X.}~\bibnamefont
			{Zhou}}, \bibinfo {author} {\bibfnamefont {Y.}~\bibnamefont {Li}}, \bibinfo
		{author} {\bibfnamefont {Z.}~\bibnamefont {Liu}}, \bibinfo {author}
		{\bibfnamefont {J.}~\bibnamefont {Hao}}, \bibinfo {author} {\bibfnamefont
			{Y.}~\bibnamefont {Dai}}, \bibinfo {author} {\bibfnamefont {Z.}~\bibnamefont
			{Wang}}, \bibinfo {author} {\bibfnamefont {Y.}~\bibnamefont {Yao}},\ and\
		\bibinfo {author} {\bibfnamefont {H.-H.}\ \bibnamefont {Wen}},\ }\bibfield
	{title} {\bibinfo {title} {Effects of niobium doping on the charge density
			wave and electronic correlations in the kagome metal
			{Cs(V$_{1-x}$Nb$_x$)$_{3}$Sb$_5$}},\ }\href
	{https://doi.org/10.1103/PhysRevB.107.125124} {\bibfield  {journal} {\bibinfo
			{journal} {Phys. Rev. B}\ }\textbf {\bibinfo {volume} {107}},\ \bibinfo
		{pages} {125124} (\bibinfo {year} {2023}{\natexlab{a}})}\BibitemShut
	{NoStop}%
	\bibitem [{\citenamefont {Ding}\ \emph {et~al.}(2022)\citenamefont {Ding},
		\citenamefont {Wo}, \citenamefont {Gu}, \citenamefont {Gu},\ and\
		\citenamefont {Zhao}}]{RN1238}%
	\BibitemOpen
	\bibfield  {author} {\bibinfo {author} {\bibfnamefont {G.}~\bibnamefont
			{Ding}}, \bibinfo {author} {\bibfnamefont {H.}~\bibnamefont {Wo}}, \bibinfo
		{author} {\bibfnamefont {Y.}~\bibnamefont {Gu}}, \bibinfo {author}
		{\bibfnamefont {Y.}~\bibnamefont {Gu}},\ and\ \bibinfo {author}
		{\bibfnamefont {J.}~\bibnamefont {Zhao}},\ }\bibfield  {title} {\bibinfo
		{title} {Effect of chromium doping on superconductivity and charge density
			wave order in the kagome metal {Cs(V$_{1-x}$Cr$_x$)$_3$Sb$_5$}},\ }\href
	{https://doi.org/10.1103/PhysRevB.106.235151} {\bibfield  {journal} {\bibinfo
			{journal} {Phys. Rev. B}\ }\textbf {\bibinfo {volume} {106}},\ \bibinfo
		{pages} {235151} (\bibinfo {year} {2022})}\BibitemShut {NoStop}%
	\bibitem [{\citenamefont {Liu}\ \emph {et~al.}(2021)\citenamefont {Liu},
		\citenamefont {Wang}, \citenamefont {Cai}, \citenamefont {Hao}, \citenamefont
		{Ma}, \citenamefont {Wang}, \citenamefont {Liu}, \citenamefont {Chen},
		\citenamefont {Zhou}, \citenamefont {Wang}, \citenamefont {Wang},
		\citenamefont {He}, \citenamefont {Liu}, \citenamefont {Cui}, \citenamefont
		{Wang}, \citenamefont {Huang}, \citenamefont {Chen},\ and\ \citenamefont
		{Mei}}]{RN202}%
	\BibitemOpen
	\bibfield  {author} {\bibinfo {author} {\bibfnamefont {Y.}~\bibnamefont
			{Liu}}, \bibinfo {author} {\bibfnamefont {Y.}~\bibnamefont {Wang}}, \bibinfo
		{author} {\bibfnamefont {Y.}~\bibnamefont {Cai}}, \bibinfo {author}
		{\bibfnamefont {Z.}~\bibnamefont {Hao}}, \bibinfo {author} {\bibfnamefont
			{X.-M.}\ \bibnamefont {Ma}}, \bibinfo {author} {\bibfnamefont
			{L.}~\bibnamefont {Wang}}, \bibinfo {author} {\bibfnamefont {C.}~\bibnamefont
			{Liu}}, \bibinfo {author} {\bibfnamefont {J.}~\bibnamefont {Chen}}, \bibinfo
		{author} {\bibfnamefont {L.}~\bibnamefont {Zhou}}, \bibinfo {author}
		{\bibfnamefont {J.}~\bibnamefont {Wang}}, \bibinfo {author} {\bibfnamefont
			{S.}~\bibnamefont {Wang}}, \bibinfo {author} {\bibfnamefont {H.}~\bibnamefont
			{He}}, \bibinfo {author} {\bibfnamefont {Y.}~\bibnamefont {Liu}}, \bibinfo
		{author} {\bibfnamefont {S.}~\bibnamefont {Cui}}, \bibinfo {author}
		{\bibfnamefont {J.}~\bibnamefont {Wang}}, \bibinfo {author} {\bibfnamefont
			{B.}~\bibnamefont {Huang}}, \bibinfo {author} {\bibfnamefont
			{C.}~\bibnamefont {Chen}},\ and\ \bibinfo {author} {\bibfnamefont {J.-W.}\
			\bibnamefont {Mei}},\ }\bibfield  {title} {\bibinfo {title} {Doping evolution
			of superconductivity, charge order and band topology in hole-doped
			topological kagome superconductors {Cs(V$_{1-x}$Ti$_{x}$)$_3$Sb$_5$}},\
	}\bibfield  {journal} {\bibinfo  {journal} {arXiv.}\ }\href
	{https://doi.org/10.48550/arXiv.2110.12651} {10.48550/arXiv.2110.12651}
	(\bibinfo {year} {2021})\BibitemShut {NoStop}%
	\bibitem [{\citenamefont {Hu}\ \emph {et~al.}(2022{\natexlab{a}})\citenamefont
		{Hu}, \citenamefont {Teicher}, \citenamefont {Ortiz}, \citenamefont {Luo},
		\citenamefont {Peng}, \citenamefont {Huai}, \citenamefont {Ma}, \citenamefont
		{Plumb}, \citenamefont {Wilson}, \citenamefont {He},\ and\ \citenamefont
		{Shi}}]{RN461}%
	\BibitemOpen
	\bibfield  {author} {\bibinfo {author} {\bibfnamefont {Y.}~\bibnamefont
			{Hu}}, \bibinfo {author} {\bibfnamefont {S.~M.~L.}\ \bibnamefont {Teicher}},
		\bibinfo {author} {\bibfnamefont {B.~R.}\ \bibnamefont {Ortiz}}, \bibinfo
		{author} {\bibfnamefont {Y.}~\bibnamefont {Luo}}, \bibinfo {author}
		{\bibfnamefont {S.}~\bibnamefont {Peng}}, \bibinfo {author} {\bibfnamefont
			{L.}~\bibnamefont {Huai}}, \bibinfo {author} {\bibfnamefont {J.}~\bibnamefont
			{Ma}}, \bibinfo {author} {\bibfnamefont {N.~C.}\ \bibnamefont {Plumb}},
		\bibinfo {author} {\bibfnamefont {S.~D.}\ \bibnamefont {Wilson}}, \bibinfo
		{author} {\bibfnamefont {J.}~\bibnamefont {He}},\ and\ \bibinfo {author}
		{\bibfnamefont {M.}~\bibnamefont {Shi}},\ }\bibfield  {title} {\bibinfo
		{title} {Topological surface states and flat bands in the kagome
			superconductor {{CsV$_3$Sb$_5$}}},\ }\href
	{https://doi.org/10.1016/j.scib.2021.11.026} {\bibfield  {journal} {\bibinfo
			{journal} {Sci. Bull.}\ }\textbf {\bibinfo {volume} {67}},\ \bibinfo {pages}
		{495} (\bibinfo {year} {2022}{\natexlab{a}})}\BibitemShut {NoStop}%
	\bibitem [{\citenamefont {Liang}\ \emph {et~al.}(2021)\citenamefont {Liang},
		\citenamefont {Hou}, \citenamefont {Zhang}, \citenamefont {Ma}, \citenamefont
		{Wu}, \citenamefont {Zhang}, \citenamefont {Yu}, \citenamefont {Ying},
		\citenamefont {Jiang}, \citenamefont {Shan}, \citenamefont {Wang},\ and\
		\citenamefont {Chen}}]{RN1}%
	\BibitemOpen
	\bibfield  {author} {\bibinfo {author} {\bibfnamefont {Z.}~\bibnamefont
			{Liang}}, \bibinfo {author} {\bibfnamefont {X.}~\bibnamefont {Hou}}, \bibinfo
		{author} {\bibfnamefont {F.}~\bibnamefont {Zhang}}, \bibinfo {author}
		{\bibfnamefont {W.}~\bibnamefont {Ma}}, \bibinfo {author} {\bibfnamefont
			{P.}~\bibnamefont {Wu}}, \bibinfo {author} {\bibfnamefont {Z.}~\bibnamefont
			{Zhang}}, \bibinfo {author} {\bibfnamefont {F.}~\bibnamefont {Yu}}, \bibinfo
		{author} {\bibfnamefont {J.~J.}\ \bibnamefont {Ying}}, \bibinfo {author}
		{\bibfnamefont {K.}~\bibnamefont {Jiang}}, \bibinfo {author} {\bibfnamefont
			{L.}~\bibnamefont {Shan}}, \bibinfo {author} {\bibfnamefont {Z.}~\bibnamefont
			{Wang}},\ and\ \bibinfo {author} {\bibfnamefont {X.~H.}\ \bibnamefont
			{Chen}},\ }\bibfield  {title} {\bibinfo {title} {Three-dimensional charge
			density wave and surface-dependent vortex-core states in a kagome
			superconductor {CsV$_3$Sb$_5$}},\ }\href
	{https://doi.org/10.1103/PhysRevX.11.031026} {\bibfield  {journal} {\bibinfo
			{journal} {Phys. Rev. X}\ }\textbf {\bibinfo {volume} {11}},\ \bibinfo
		{pages} {031026} (\bibinfo {year} {2021})}\BibitemShut {NoStop}%
	\bibitem [{\citenamefont {Yi}\ \emph {et~al.}(2022)\citenamefont {Yi},
		\citenamefont {Ma}, \citenamefont {Zhang}, \citenamefont {Liao},
		\citenamefont {You},\ and\ \citenamefont {Su}}]{RN985}%
	\BibitemOpen
	\bibfield  {author} {\bibinfo {author} {\bibfnamefont {X.-W.}\ \bibnamefont
			{Yi}}, \bibinfo {author} {\bibfnamefont {X.-Y.}\ \bibnamefont {Ma}}, \bibinfo
		{author} {\bibfnamefont {Z.}~\bibnamefont {Zhang}}, \bibinfo {author}
		{\bibfnamefont {Z.-W.}\ \bibnamefont {Liao}}, \bibinfo {author}
		{\bibfnamefont {J.-Y.}\ \bibnamefont {You}},\ and\ \bibinfo {author}
		{\bibfnamefont {G.}~\bibnamefont {Su}},\ }\bibfield  {title} {\bibinfo
		{title} {Large kagome family candidates with topological superconductivity
			and charge density waves},\ }\href
	{https://doi.org/10.1103/PhysRevB.106.L220505} {\bibfield  {journal}
		{\bibinfo  {journal} {Phys. Rev. B}\ }\textbf {\bibinfo {volume} {106}},\
		\bibinfo {pages} {L220505} (\bibinfo {year} {2022})}\BibitemShut {NoStop}%
	\bibitem [{\citenamefont {Yang}\ \emph
		{et~al.}(2022{\natexlab{b}})\citenamefont {Yang}, \citenamefont {Zhao},
		\citenamefont {Yi}, \citenamefont {Liu}, \citenamefont {You}, \citenamefont
		{Zhang}, \citenamefont {Guo}, \citenamefont {Lin}, \citenamefont {Shen},
		\citenamefont {Chen}, \citenamefont {Dong}, \citenamefont {Su},\ and\
		\citenamefont {Gao}}]{RN565}%
	\BibitemOpen
	\bibfield  {author} {\bibinfo {author} {\bibfnamefont {H.}~\bibnamefont
			{Yang}}, \bibinfo {author} {\bibfnamefont {Z.}~\bibnamefont {Zhao}}, \bibinfo
		{author} {\bibfnamefont {X.-W.}\ \bibnamefont {Yi}}, \bibinfo {author}
		{\bibfnamefont {J.}~\bibnamefont {Liu}}, \bibinfo {author} {\bibfnamefont
			{J.-Y.}\ \bibnamefont {You}}, \bibinfo {author} {\bibfnamefont
			{Y.}~\bibnamefont {Zhang}}, \bibinfo {author} {\bibfnamefont
			{H.}~\bibnamefont {Guo}}, \bibinfo {author} {\bibfnamefont {X.}~\bibnamefont
			{Lin}}, \bibinfo {author} {\bibfnamefont {C.}~\bibnamefont {Shen}}, \bibinfo
		{author} {\bibfnamefont {H.}~\bibnamefont {Chen}}, \bibinfo {author}
		{\bibfnamefont {X.}~\bibnamefont {Dong}}, \bibinfo {author} {\bibfnamefont
			{G.}~\bibnamefont {Su}},\ and\ \bibinfo {author} {\bibfnamefont {H.-J.}\
			\bibnamefont {Gao}},\ }\bibfield  {title} {\bibinfo {title} {Titanium-based
			kagome superconductor {CsTi$_{3}$Bi$_{5}$} and topological states},\
	}\bibfield  {journal} {\bibinfo  {journal} {arXiv.}\ }\href
	{https://doi.org/10.48550/arXiv.2209.03840} {10.48550/arXiv.2209.03840}
	(\bibinfo {year} {2022}{\natexlab{b}})\BibitemShut {NoStop}%
	\bibitem [{\citenamefont {Yang}\ \emph
		{et~al.}(2022{\natexlab{c}})\citenamefont {Yang}, \citenamefont {Ye},
		\citenamefont {Zhao}, \citenamefont {Liu}, \citenamefont {Yi}, \citenamefont
		{Zhang}, \citenamefont {Shi}, \citenamefont {You}, \citenamefont {Huang},
		\citenamefont {Wang}, \citenamefont {Wang}, \citenamefont {Guo},
		\citenamefont {Lin}, \citenamefont {Shen}, \citenamefont {Zhou},
		\citenamefont {Chen}, \citenamefont {Dong}, \citenamefont {Su}, \citenamefont
		{Wang},\ and\ \citenamefont {Gao}}]{RN967}%
	\BibitemOpen
	\bibfield  {author} {\bibinfo {author} {\bibfnamefont {H.}~\bibnamefont
			{Yang}}, \bibinfo {author} {\bibfnamefont {Y.}~\bibnamefont {Ye}}, \bibinfo
		{author} {\bibfnamefont {Z.}~\bibnamefont {Zhao}}, \bibinfo {author}
		{\bibfnamefont {J.}~\bibnamefont {Liu}}, \bibinfo {author} {\bibfnamefont
			{X.-W.}\ \bibnamefont {Yi}}, \bibinfo {author} {\bibfnamefont
			{Y.}~\bibnamefont {Zhang}}, \bibinfo {author} {\bibfnamefont
			{J.}~\bibnamefont {Shi}}, \bibinfo {author} {\bibfnamefont {J.-Y.}\
			\bibnamefont {You}}, \bibinfo {author} {\bibfnamefont {Z.}~\bibnamefont
			{Huang}}, \bibinfo {author} {\bibfnamefont {B.}~\bibnamefont {Wang}},
		\bibinfo {author} {\bibfnamefont {J.}~\bibnamefont {Wang}}, \bibinfo {author}
		{\bibfnamefont {H.}~\bibnamefont {Guo}}, \bibinfo {author} {\bibfnamefont
			{X.}~\bibnamefont {Lin}}, \bibinfo {author} {\bibfnamefont {C.}~\bibnamefont
			{Shen}}, \bibinfo {author} {\bibfnamefont {W.}~\bibnamefont {Zhou}}, \bibinfo
		{author} {\bibfnamefont {H.}~\bibnamefont {Chen}}, \bibinfo {author}
		{\bibfnamefont {X.}~\bibnamefont {Dong}}, \bibinfo {author} {\bibfnamefont
			{G.}~\bibnamefont {Su}}, \bibinfo {author} {\bibfnamefont {Z.}~\bibnamefont
			{Wang}},\ and\ \bibinfo {author} {\bibfnamefont {H.-J.}\ \bibnamefont
			{Gao}},\ }\bibfield  {title} {\bibinfo {title} {Superconductivity and
			orbital-selective nematic order in a new titaniumbased kagome metal
			{CsTi$_{3}$Bi$_{5}$}},\ }\bibfield  {journal} {\bibinfo  {journal} {arXiv.}\
	}\href {https://doi.org/10.48550/arXiv.2211.12264}
	{10.48550/arXiv.2211.12264} (\bibinfo {year}
	{2022}{\natexlab{c}})\BibitemShut {NoStop}%
	\bibitem [{\citenamefont {Werhahn}\ \emph {et~al.}(2022)\citenamefont
		{Werhahn}, \citenamefont {Ortiz}, \citenamefont {Hay}, \citenamefont
		{Wilson}, \citenamefont {Seshadri},\ and\ \citenamefont {Johrendt}}]{RN795}%
	\BibitemOpen
	\bibfield  {author} {\bibinfo {author} {\bibfnamefont {D.}~\bibnamefont
			{Werhahn}}, \bibinfo {author} {\bibfnamefont {B.~R.}\ \bibnamefont {Ortiz}},
		\bibinfo {author} {\bibfnamefont {A.~K.}\ \bibnamefont {Hay}}, \bibinfo
		{author} {\bibfnamefont {S.~D.}\ \bibnamefont {Wilson}}, \bibinfo {author}
		{\bibfnamefont {R.}~\bibnamefont {Seshadri}},\ and\ \bibinfo {author}
		{\bibfnamefont {D.}~\bibnamefont {Johrendt}},\ }\bibfield  {title} {\bibinfo
		{title} {The kagomé metals {RbTi$_{3}$Bi$_{5}$} and {CsTi$_{3}$Bi$_{5}$}},\
	}\href {https://doi.org/10.1515/znb-2022-0125} {\bibfield  {journal}
		{\bibinfo  {journal} {Zeitschrift für Naturforschung B}\ }\textbf {\bibinfo
			{volume} {77}},\ \bibinfo {pages} {757} (\bibinfo {year} {2022})}\BibitemShut
	{NoStop}%
	\bibitem [{\citenamefont {Li}\ \emph {et~al.}(2022)\citenamefont {Li},
		\citenamefont {Cheng}, \citenamefont {Ortiz}, \citenamefont {Tan},
		\citenamefont {Werhahn}, \citenamefont {Keyu}, \citenamefont {Jorhendt},
		\citenamefont {Yan}, \citenamefont {Wang}, \citenamefont {Wilson},\ and\
		\citenamefont {Zeljkovic}}]{RN955}%
	\BibitemOpen
	\bibfield  {author} {\bibinfo {author} {\bibfnamefont {H.}~\bibnamefont
			{Li}}, \bibinfo {author} {\bibfnamefont {S.}~\bibnamefont {Cheng}}, \bibinfo
		{author} {\bibfnamefont {B.~R.}\ \bibnamefont {Ortiz}}, \bibinfo {author}
		{\bibfnamefont {H.}~\bibnamefont {Tan}}, \bibinfo {author} {\bibfnamefont
			{D.}~\bibnamefont {Werhahn}}, \bibinfo {author} {\bibfnamefont
			{Z.}~\bibnamefont {Keyu}}, \bibinfo {author} {\bibfnamefont {D.}~\bibnamefont
			{Jorhendt}}, \bibinfo {author} {\bibfnamefont {B.}~\bibnamefont {Yan}},
		\bibinfo {author} {\bibfnamefont {Z.}~\bibnamefont {Wang}}, \bibinfo {author}
		{\bibfnamefont {S.~D.}\ \bibnamefont {Wilson}},\ and\ \bibinfo {author}
		{\bibfnamefont {I.}~\bibnamefont {Zeljkovic}},\ }\bibfield  {title} {\bibinfo
		{title} {Electronic nematicity in the absence of charge density waves in a
			new titanium-based kagome metal},\ }\bibfield  {journal} {\bibinfo  {journal}
		{arXiv.}\ }\href {https://doi.org/10.48550/arXiv.2211.16477}
	{10.48550/arXiv.2211.16477} (\bibinfo {year} {2022})\BibitemShut {NoStop}%
	\bibitem [{\citenamefont {Yi}\ \emph {et~al.}(2023)\citenamefont {Yi},
		\citenamefont {Liao}, \citenamefont {You}, \citenamefont {Gu},\ and\
		\citenamefont {Su}}]{RN1006}%
	\BibitemOpen
	\bibfield  {author} {\bibinfo {author} {\bibfnamefont {X.~W.}\ \bibnamefont
			{Yi}}, \bibinfo {author} {\bibfnamefont {Z.~W.}\ \bibnamefont {Liao}},
		\bibinfo {author} {\bibfnamefont {J.~Y.}\ \bibnamefont {You}}, \bibinfo
		{author} {\bibfnamefont {B.}~\bibnamefont {Gu}},\ and\ \bibinfo {author}
		{\bibfnamefont {G.}~\bibnamefont {Su}},\ }\bibfield  {title} {\bibinfo
		{title} {Topological superconductivity and large spin {Hall} effect in the
			kagome family {Ti$_6$X$_4$ (X = Bi, Sb, Pb, Tl, and In)}},\ }\href
	{https://doi.org/10.1016/j.isci.2022.105813} {\bibfield  {journal} {\bibinfo
			{journal} {iScience}\ }\textbf {\bibinfo {volume} {26}},\ \bibinfo {pages}
		{105813} (\bibinfo {year} {2023})}\BibitemShut {NoStop}%
	\bibitem [{\citenamefont {Chen}\ \emph {et~al.}(2023)\citenamefont {Chen},
		\citenamefont {Liu}, \citenamefont {Xia}, \citenamefont {Mi}, \citenamefont
		{Zhong}, \citenamefont {Yang}, \citenamefont {Zhang}, \citenamefont {Gan},
		\citenamefont {Liu}, \citenamefont {Wang}, \citenamefont {Wang},
		\citenamefont {Chai}, \citenamefont {Shen}, \citenamefont {Yang},
		\citenamefont {Guo},\ and\ \citenamefont {He}}]{RN1226}%
	\BibitemOpen
	\bibfield  {author} {\bibinfo {author} {\bibfnamefont {X.}~\bibnamefont
			{Chen}}, \bibinfo {author} {\bibfnamefont {X.}~\bibnamefont {Liu}}, \bibinfo
		{author} {\bibfnamefont {W.}~\bibnamefont {Xia}}, \bibinfo {author}
		{\bibfnamefont {X.}~\bibnamefont {Mi}}, \bibinfo {author} {\bibfnamefont
			{L.}~\bibnamefont {Zhong}}, \bibinfo {author} {\bibfnamefont
			{K.}~\bibnamefont {Yang}}, \bibinfo {author} {\bibfnamefont {L.}~\bibnamefont
			{Zhang}}, \bibinfo {author} {\bibfnamefont {Y.}~\bibnamefont {Gan}}, \bibinfo
		{author} {\bibfnamefont {Y.}~\bibnamefont {Liu}}, \bibinfo {author}
		{\bibfnamefont {G.}~\bibnamefont {Wang}}, \bibinfo {author} {\bibfnamefont
			{A.}~\bibnamefont {Wang}}, \bibinfo {author} {\bibfnamefont {Y.}~\bibnamefont
			{Chai}}, \bibinfo {author} {\bibfnamefont {J.}~\bibnamefont {Shen}}, \bibinfo
		{author} {\bibfnamefont {X.}~\bibnamefont {Yang}}, \bibinfo {author}
		{\bibfnamefont {Y.}~\bibnamefont {Guo}},\ and\ \bibinfo {author}
		{\bibfnamefont {M.}~\bibnamefont {He}},\ }\bibfield  {title} {\bibinfo
		{title} {Electrical and thermal transport properties of the kagome metals
			{ATi$_3$Bi$_5$(A=Rb,Cs)}},\ }\href
	{https://doi.org/10.1103/PhysRevB.107.174510} {\bibfield  {journal} {\bibinfo
			{journal} {Phys. Rev. B}\ }\textbf {\bibinfo {volume} {107}},\ \bibinfo
		{pages} {174510} (\bibinfo {year} {2023})}\BibitemShut {NoStop}%
	\bibitem [{\citenamefont {Yang}\ \emph
		{et~al.}(2022{\natexlab{d}})\citenamefont {Yang}, \citenamefont {Yi},
		\citenamefont {Zhao}, \citenamefont {Xie}, \citenamefont {Miao},
		\citenamefont {Luo}, \citenamefont {Chen}, \citenamefont {Liang},
		\citenamefont {Zhu}, \citenamefont {Ye}, \citenamefont {You}, \citenamefont
		{Gu}, \citenamefont {Zhang}, \citenamefont {Zhang}, \citenamefont {Yang},
		\citenamefont {Wang}, \citenamefont {Peng}, \citenamefont {Mao},
		\citenamefont {Liu}, \citenamefont {Xu}, \citenamefont {Chen}, \citenamefont
		{Yang}, \citenamefont {Su}, \citenamefont {Gao}, \citenamefont {Zhao},\ and\
		\citenamefont {Zhou}}]{RN966}%
	\BibitemOpen
	\bibfield  {author} {\bibinfo {author} {\bibfnamefont {J.}~\bibnamefont
			{Yang}}, \bibinfo {author} {\bibfnamefont {X.}~\bibnamefont {Yi}}, \bibinfo
		{author} {\bibfnamefont {Z.}~\bibnamefont {Zhao}}, \bibinfo {author}
		{\bibfnamefont {Y.}~\bibnamefont {Xie}}, \bibinfo {author} {\bibfnamefont
			{T.}~\bibnamefont {Miao}}, \bibinfo {author} {\bibfnamefont {H.}~\bibnamefont
			{Luo}}, \bibinfo {author} {\bibfnamefont {H.}~\bibnamefont {Chen}}, \bibinfo
		{author} {\bibfnamefont {B.}~\bibnamefont {Liang}}, \bibinfo {author}
		{\bibfnamefont {W.}~\bibnamefont {Zhu}}, \bibinfo {author} {\bibfnamefont
			{Y.}~\bibnamefont {Ye}}, \bibinfo {author} {\bibfnamefont {J.-Y.}\
			\bibnamefont {You}}, \bibinfo {author} {\bibfnamefont {B.}~\bibnamefont
			{Gu}}, \bibinfo {author} {\bibfnamefont {S.}~\bibnamefont {Zhang}}, \bibinfo
		{author} {\bibfnamefont {F.}~\bibnamefont {Zhang}}, \bibinfo {author}
		{\bibfnamefont {F.}~\bibnamefont {Yang}}, \bibinfo {author} {\bibfnamefont
			{Z.}~\bibnamefont {Wang}}, \bibinfo {author} {\bibfnamefont {Q.}~\bibnamefont
			{Peng}}, \bibinfo {author} {\bibfnamefont {H.}~\bibnamefont {Mao}}, \bibinfo
		{author} {\bibfnamefont {G.}~\bibnamefont {Liu}}, \bibinfo {author}
		{\bibfnamefont {Z.}~\bibnamefont {Xu}}, \bibinfo {author} {\bibfnamefont
			{H.}~\bibnamefont {Chen}}, \bibinfo {author} {\bibfnamefont {H.}~\bibnamefont
			{Yang}}, \bibinfo {author} {\bibfnamefont {G.}~\bibnamefont {Su}}, \bibinfo
		{author} {\bibfnamefont {H.}~\bibnamefont {Gao}}, \bibinfo {author}
		{\bibfnamefont {L.}~\bibnamefont {Zhao}},\ and\ \bibinfo {author}
		{\bibfnamefont {X.~J.}\ \bibnamefont {Zhou}},\ }\bibfield  {title} {\bibinfo
		{title} {Observation of flat band, {Dirac} nodal lines and topological
			surface states in kagome superconductor {CsTi$_{3}$Bi$_{5}$}},\ }\bibfield
	{journal} {\bibinfo  {journal} {Nat. Commun. accepted}\ }\href
	{https://doi.org/10.48550/arXiv.2212.04447} {10.48550/arXiv.2212.04447}
	(\bibinfo {year} {2022}{\natexlab{d}})\BibitemShut {NoStop}%
	\bibitem [{\citenamefont {Hu}\ \emph {et~al.}(2022{\natexlab{b}})\citenamefont
		{Hu}, \citenamefont {Le}, \citenamefont {Zhao}, \citenamefont {Ma},
		\citenamefont {Plumb}, \citenamefont {Radovic}, \citenamefont {Schnyder},
		\citenamefont {Wu}, \citenamefont {Chen}, \citenamefont {Dong}, \citenamefont
		{Hu}, \citenamefont {Yang}, \citenamefont {Gao},\ and\ \citenamefont
		{Shi}}]{RN970}%
	\BibitemOpen
	\bibfield  {author} {\bibinfo {author} {\bibfnamefont {Y.}~\bibnamefont
			{Hu}}, \bibinfo {author} {\bibfnamefont {C.}~\bibnamefont {Le}}, \bibinfo
		{author} {\bibfnamefont {Z.}~\bibnamefont {Zhao}}, \bibinfo {author}
		{\bibfnamefont {J.}~\bibnamefont {Ma}}, \bibinfo {author} {\bibfnamefont
			{N.~C.}\ \bibnamefont {Plumb}}, \bibinfo {author} {\bibfnamefont
			{M.}~\bibnamefont {Radovic}}, \bibinfo {author} {\bibfnamefont {A.~P.}\
			\bibnamefont {Schnyder}}, \bibinfo {author} {\bibfnamefont {X.}~\bibnamefont
			{Wu}}, \bibinfo {author} {\bibfnamefont {H.}~\bibnamefont {Chen}}, \bibinfo
		{author} {\bibfnamefont {X.}~\bibnamefont {Dong}}, \bibinfo {author}
		{\bibfnamefont {J.}~\bibnamefont {Hu}}, \bibinfo {author} {\bibfnamefont
			{H.}~\bibnamefont {Yang}}, \bibinfo {author} {\bibfnamefont {H.-J.}\
			\bibnamefont {Gao}},\ and\ \bibinfo {author} {\bibfnamefont {M.}~\bibnamefont
			{Shi}},\ }\bibfield  {title} {\bibinfo {title} {Non-trivial band topology and
			orbital-selective electronic nematicity in a new titanium-based kagome
			superconductor},\ }\bibfield  {journal} {\bibinfo  {journal} {arXiv.}\ }\href
	{https://doi.org/10.48550/arXiv.2212.07958} {10.48550/arXiv.2212.07958}
	(\bibinfo {year} {2022}{\natexlab{b}})\BibitemShut {NoStop}%
	\bibitem [{\citenamefont {Jiang}\ \emph {et~al.}(2022)\citenamefont {Jiang},
		\citenamefont {Liu}, \citenamefont {Ma}, \citenamefont {Xia}, \citenamefont
		{Liu}, \citenamefont {Liu}, \citenamefont {Cho}, \citenamefont {Yang},
		\citenamefont {Ding}, \citenamefont {Liu}, \citenamefont {Huang},
		\citenamefont {Qiao}, \citenamefont {Shen}, \citenamefont {Jing},
		\citenamefont {Liu}, \citenamefont {Liu}, \citenamefont {Guo},\ and\
		\citenamefont {Shen}}]{RN968}%
	\BibitemOpen
	\bibfield  {author} {\bibinfo {author} {\bibfnamefont {Z.}~\bibnamefont
			{Jiang}}, \bibinfo {author} {\bibfnamefont {Z.}~\bibnamefont {Liu}}, \bibinfo
		{author} {\bibfnamefont {H.}~\bibnamefont {Ma}}, \bibinfo {author}
		{\bibfnamefont {W.}~\bibnamefont {Xia}}, \bibinfo {author} {\bibfnamefont
			{Z.}~\bibnamefont {Liu}}, \bibinfo {author} {\bibfnamefont {J.}~\bibnamefont
			{Liu}}, \bibinfo {author} {\bibfnamefont {S.}~\bibnamefont {Cho}}, \bibinfo
		{author} {\bibfnamefont {Y.}~\bibnamefont {Yang}}, \bibinfo {author}
		{\bibfnamefont {J.}~\bibnamefont {Ding}}, \bibinfo {author} {\bibfnamefont
			{J.}~\bibnamefont {Liu}}, \bibinfo {author} {\bibfnamefont {Z.}~\bibnamefont
			{Huang}}, \bibinfo {author} {\bibfnamefont {Y.}~\bibnamefont {Qiao}},
		\bibinfo {author} {\bibfnamefont {J.}~\bibnamefont {Shen}}, \bibinfo {author}
		{\bibfnamefont {W.}~\bibnamefont {Jing}}, \bibinfo {author} {\bibfnamefont
			{X.}~\bibnamefont {Liu}}, \bibinfo {author} {\bibfnamefont {J.}~\bibnamefont
			{Liu}}, \bibinfo {author} {\bibfnamefont {Y.}~\bibnamefont {Guo}},\ and\
		\bibinfo {author} {\bibfnamefont {D.}~\bibnamefont {Shen}},\ }\bibfield
	{title} {\bibinfo {title} {Flat bands, non-trivial band topology and
			electronic nematicity in layered kagome-lattice {RbTi$_{3}$Bi$_{5}$}},\
	}\bibfield  {journal} {\bibinfo  {journal} {arXiv.}\ }\href
	{https://doi.org/10.48550/arXiv.2212.02399} {10.48550/arXiv.2212.02399}
	(\bibinfo {year} {2022})\BibitemShut {NoStop}%
	\bibitem [{\citenamefont {Zhou}\ \emph
		{et~al.}(2023{\natexlab{b}})\citenamefont {Zhou}, \citenamefont {Chen},
		\citenamefont {Ji}, \citenamefont {Liu}, \citenamefont {Liao}, \citenamefont
		{Guo}, \citenamefont {Wang}, \citenamefont {Weng},\ and\ \citenamefont
		{Wang}}]{RN1018}%
	\BibitemOpen
	\bibfield  {author} {\bibinfo {author} {\bibfnamefont {Y.}~\bibnamefont
			{Zhou}}, \bibinfo {author} {\bibfnamefont {L.}~\bibnamefont {Chen}}, \bibinfo
		{author} {\bibfnamefont {X.}~\bibnamefont {Ji}}, \bibinfo {author}
		{\bibfnamefont {C.}~\bibnamefont {Liu}}, \bibinfo {author} {\bibfnamefont
			{K.}~\bibnamefont {Liao}}, \bibinfo {author} {\bibfnamefont {Z.}~\bibnamefont
			{Guo}}, \bibinfo {author} {\bibfnamefont {J.~o.}\ \bibnamefont {Wang}},
		\bibinfo {author} {\bibfnamefont {H.}~\bibnamefont {Weng}},\ and\ \bibinfo
		{author} {\bibfnamefont {G.}~\bibnamefont {Wang}},\ }\bibfield  {title}
	{\bibinfo {title} {Physical properties, electronic structure, and
			strain-tuned monolayer of the weak topological insulator {RbTi$_{3}$Bi$_{5}$}
			with kagome lattice},\ }\bibfield  {journal} {\bibinfo  {journal} {arXiv.}\
	}\href {https://doi.org/10.48550/arXiv.2301.01633}
	{10.48550/arXiv.2301.01633} (\bibinfo {year}
	{2023}{\natexlab{b}})\BibitemShut {NoStop}%
	\bibitem [{\citenamefont {Liu}\ \emph {et~al.}(2022{\natexlab{c}})\citenamefont
		{Liu}, \citenamefont {Kuang}, \citenamefont {Luo}, \citenamefont {Li},
		\citenamefont {Huai}, \citenamefont {Peng}, \citenamefont {Wei},
		\citenamefont {Shen}, \citenamefont {Wang}, \citenamefont {Miao},
		\citenamefont {Sun}, \citenamefont {Ou}, \citenamefont {Yao}, \citenamefont
		{Wang},\ and\ \citenamefont {He}}]{RN965}%
	\BibitemOpen
	\bibfield  {author} {\bibinfo {author} {\bibfnamefont {B.}~\bibnamefont
			{Liu}}, \bibinfo {author} {\bibfnamefont {M.}~\bibnamefont {Kuang}}, \bibinfo
		{author} {\bibfnamefont {Y.}~\bibnamefont {Luo}}, \bibinfo {author}
		{\bibfnamefont {Y.}~\bibnamefont {Li}}, \bibinfo {author} {\bibfnamefont
			{L.}~\bibnamefont {Huai}}, \bibinfo {author} {\bibfnamefont {S.}~\bibnamefont
			{Peng}}, \bibinfo {author} {\bibfnamefont {Z.}~\bibnamefont {Wei}}, \bibinfo
		{author} {\bibfnamefont {J.}~\bibnamefont {Shen}}, \bibinfo {author}
		{\bibfnamefont {B.}~\bibnamefont {Wang}}, \bibinfo {author} {\bibfnamefont
			{Y.}~\bibnamefont {Miao}}, \bibinfo {author} {\bibfnamefont {X.}~\bibnamefont
			{Sun}}, \bibinfo {author} {\bibfnamefont {Z.}~\bibnamefont {Ou}}, \bibinfo
		{author} {\bibfnamefont {Y.}~\bibnamefont {Yao}}, \bibinfo {author}
		{\bibfnamefont {Z.}~\bibnamefont {Wang}},\ and\ \bibinfo {author}
		{\bibfnamefont {J.}~\bibnamefont {He}},\ }\bibfield  {title} {\bibinfo
		{title} {Tunable van {Hove} singularity without structural instability in
			kagome metal {CsTi$_{3}$Bi$_{5}$}},\ }\bibfield  {journal} {\bibinfo
		{journal} {arXiv.}\ }\href {https://doi.org/10.48550/arXiv.2212.04460}
	{10.48550/arXiv.2212.04460} (\bibinfo {year}
	{2022}{\natexlab{c}})\BibitemShut {NoStop}%
	\bibitem [{\citenamefont {Wang}\ \emph {et~al.}(2023)\citenamefont {Wang},
		\citenamefont {Liu}, \citenamefont {Hao}, \citenamefont {Cheng},
		\citenamefont {Deng}, \citenamefont {Wang}, \citenamefont {Gu}, \citenamefont
		{Ma}, \citenamefont {Rong}, \citenamefont {Zhang}, \citenamefont {Guo},
		\citenamefont {Zhang}, \citenamefont {Jiang}, \citenamefont {Yang},
		\citenamefont {Liu}, \citenamefont {Jiang}, \citenamefont {Liu},
		\citenamefont {Ye}, \citenamefont {Shen}, \citenamefont {Liu}, \citenamefont
		{Cui}, \citenamefont {Wang}, \citenamefont {Liu}, \citenamefont {Lin},
		\citenamefont {Liu}, \citenamefont {Cai}, \citenamefont {Zhu}, \citenamefont
		{Chen},\ and\ \citenamefont {Mei}}]{RN1123}%
	\BibitemOpen
	\bibfield  {author} {\bibinfo {author} {\bibfnamefont {Y.}~\bibnamefont
			{Wang}}, \bibinfo {author} {\bibfnamefont {Y.}~\bibnamefont {Liu}}, \bibinfo
		{author} {\bibfnamefont {Z.}~\bibnamefont {Hao}}, \bibinfo {author}
		{\bibfnamefont {W.}~\bibnamefont {Cheng}}, \bibinfo {author} {\bibfnamefont
			{J.}~\bibnamefont {Deng}}, \bibinfo {author} {\bibfnamefont {Y.}~\bibnamefont
			{Wang}}, \bibinfo {author} {\bibfnamefont {Y.}~\bibnamefont {Gu}}, \bibinfo
		{author} {\bibfnamefont {X.-M.}\ \bibnamefont {Ma}}, \bibinfo {author}
		{\bibfnamefont {H.}~\bibnamefont {Rong}}, \bibinfo {author} {\bibfnamefont
			{F.}~\bibnamefont {Zhang}}, \bibinfo {author} {\bibfnamefont
			{S.}~\bibnamefont {Guo}}, \bibinfo {author} {\bibfnamefont {C.}~\bibnamefont
			{Zhang}}, \bibinfo {author} {\bibfnamefont {Z.}~\bibnamefont {Jiang}},
		\bibinfo {author} {\bibfnamefont {Y.}~\bibnamefont {Yang}}, \bibinfo {author}
		{\bibfnamefont {W.}~\bibnamefont {Liu}}, \bibinfo {author} {\bibfnamefont
			{Q.}~\bibnamefont {Jiang}}, \bibinfo {author} {\bibfnamefont
			{Z.}~\bibnamefont {Liu}}, \bibinfo {author} {\bibfnamefont {M.}~\bibnamefont
			{Ye}}, \bibinfo {author} {\bibfnamefont {D.}~\bibnamefont {Shen}}, \bibinfo
		{author} {\bibfnamefont {Y.}~\bibnamefont {Liu}}, \bibinfo {author}
		{\bibfnamefont {S.}~\bibnamefont {Cui}}, \bibinfo {author} {\bibfnamefont
			{L.}~\bibnamefont {Wang}}, \bibinfo {author} {\bibfnamefont {C.}~\bibnamefont
			{Liu}}, \bibinfo {author} {\bibfnamefont {J.}~\bibnamefont {Lin}}, \bibinfo
		{author} {\bibfnamefont {Y.}~\bibnamefont {Liu}}, \bibinfo {author}
		{\bibfnamefont {Y.}~\bibnamefont {Cai}}, \bibinfo {author} {\bibfnamefont
			{J.}~\bibnamefont {Zhu}}, \bibinfo {author} {\bibfnamefont {C.}~\bibnamefont
			{Chen}},\ and\ \bibinfo {author} {\bibfnamefont {J.-W.}\ \bibnamefont
			{Mei}},\ }\bibfield  {title} {\bibinfo {title} {Flat band and {Z$_{2}$}
			topology of kagome metal {CsTi$_{3}$Bi$_{5}$}},\ }\href
	{https://doi.org/10.1088/0256-307x/40/3/037102} {\bibfield  {journal}
		{\bibinfo  {journal} {Chin. Phys. Lett.}\ }\textbf {\bibinfo {volume} {40}},\
		\bibinfo {pages} {037102} (\bibinfo {year} {2023})}\BibitemShut {NoStop}%
	\bibitem [{\citenamefont {You}\ \emph {et~al.}(2023)\citenamefont {You},
		\citenamefont {Su},\ and\ \citenamefont {Feng}}]{RN1206}%
	\BibitemOpen
	\bibfield  {author} {\bibinfo {author} {\bibfnamefont {J.-Y.}\ \bibnamefont
			{You}}, \bibinfo {author} {\bibfnamefont {G.}~\bibnamefont {Su}},\ and\
		\bibinfo {author} {\bibfnamefont {Y.~P.}\ \bibnamefont {Feng}},\ }\bibfield
	{title} {\bibinfo {title} {A versatile model with three-dimensional
			triangular lattice for unconventional transport and various topological
			effects},\ }\bibfield  {journal} {\bibinfo  {journal} {Nat. Sci. Rev.}\
	}\href {https://doi.org/10.1093/nsr/nwad114} {10.1093/nsr/nwad114} (\bibinfo
	{year} {2023})\BibitemShut {NoStop}%
	\bibitem [{\citenamefont {Zhang}\ \emph
		{et~al.}(2021{\natexlab{b}})\citenamefont {Zhang}, \citenamefont {Liu},\ and\
		\citenamefont {Lu}}]{RN150}%
	\BibitemOpen
	\bibfield  {author} {\bibinfo {author} {\bibfnamefont {J.-F.}\ \bibnamefont
			{Zhang}}, \bibinfo {author} {\bibfnamefont {K.}~\bibnamefont {Liu}},\ and\
		\bibinfo {author} {\bibfnamefont {Z.-Y.}\ \bibnamefont {Lu}},\ }\bibfield
	{title} {\bibinfo {title} {First-principles study of the double-dome
			superconductivity in the kagome material {{CsV$_{3}$Sb$_{5}$}} under
			pressure},\ }\href {https://doi.org/10.1103/PhysRevB.104.195130} {\bibfield
		{journal} {\bibinfo  {journal} {Phys. Rev. B}\ }\textbf {\bibinfo {volume}
			{104}},\ \bibinfo {pages} {195130} (\bibinfo {year}
		{2021}{\natexlab{b}})}\BibitemShut {NoStop}%
	\bibitem [{\citenamefont {Si}\ \emph {et~al.}(2022)\citenamefont {Si},
		\citenamefont {Lu}, \citenamefont {Sun}, \citenamefont {Liu},\ and\
		\citenamefont {Wang}}]{RN193}%
	\BibitemOpen
	\bibfield  {author} {\bibinfo {author} {\bibfnamefont {J.}~\bibnamefont
			{Si}}, \bibinfo {author} {\bibfnamefont {W.}~\bibnamefont {Lu}}, \bibinfo
		{author} {\bibfnamefont {Y.}~\bibnamefont {Sun}}, \bibinfo {author}
		{\bibfnamefont {P.}~\bibnamefont {Liu}},\ and\ \bibinfo {author}
		{\bibfnamefont {B.}~\bibnamefont {Wang}},\ }\bibfield  {title} {\bibinfo
		{title} {Charge density wave and pressure-dependent superconductivity in the
			kagome metal {CsV$_3$Sb$_5$}: A first-principles study},\ }\href
	{https://doi.org/10.1103/PhysRevB.105.024517} {\bibfield  {journal} {\bibinfo
			{journal} {Phys. Rev. B}\ }\textbf {\bibinfo {volume} {105}},\ \bibinfo
		{pages} {024517} (\bibinfo {year} {2022})}\BibitemShut {NoStop}%
	\bibitem [{\citenamefont {McMillan}(1968)}]{RN128}%
	\BibitemOpen
	\bibfield  {author} {\bibinfo {author} {\bibfnamefont {W.~L.}\ \bibnamefont
			{McMillan}},\ }\bibfield  {title} {\bibinfo {title} {Transition temperature
			of strong-coupled superconductors},\ }\href
	{https://doi.org/10.1103/PhysRev.167.331} {\bibfield  {journal} {\bibinfo
			{journal} {Phys. Rev.}\ }\textbf {\bibinfo {volume} {167}},\ \bibinfo {pages}
		{331} (\bibinfo {year} {1968})}\BibitemShut {NoStop}%
	\bibitem [{\citenamefont {Allen}\ and\ \citenamefont {Dynes}(1975)}]{RN129}%
	\BibitemOpen
	\bibfield  {author} {\bibinfo {author} {\bibfnamefont {P.~B.}\ \bibnamefont
			{Allen}}\ and\ \bibinfo {author} {\bibfnamefont {R.~C.}\ \bibnamefont
			{Dynes}},\ }\bibfield  {title} {\bibinfo {title} {Transition temperature of
			strong-coupled superconductors reanalyzed},\ }\href
	{https://doi.org/10.1103/PhysRevB.12.905} {\bibfield  {journal} {\bibinfo
			{journal} {Phys. Rev. B}\ }\textbf {\bibinfo {volume} {12}},\ \bibinfo
		{pages} {905} (\bibinfo {year} {1975})}\BibitemShut {NoStop}%
	\bibitem [{\citenamefont {Tsirlin}\ \emph {et~al.}(2022)\citenamefont
		{Tsirlin}, \citenamefont {Fertey}, \citenamefont {Ortiz}, \citenamefont
		{Klis}, \citenamefont {Merkl}, \citenamefont {Dressel}, \citenamefont
		{Wilson},\ and\ \citenamefont {Uykur}}]{RN136}%
	\BibitemOpen
	\bibfield  {author} {\bibinfo {author} {\bibfnamefont {A.}~\bibnamefont
			{Tsirlin}}, \bibinfo {author} {\bibfnamefont {P.}~\bibnamefont {Fertey}},
		\bibinfo {author} {\bibfnamefont {B.~R.}\ \bibnamefont {Ortiz}}, \bibinfo
		{author} {\bibfnamefont {B.}~\bibnamefont {Klis}}, \bibinfo {author}
		{\bibfnamefont {V.}~\bibnamefont {Merkl}}, \bibinfo {author} {\bibfnamefont
			{M.}~\bibnamefont {Dressel}}, \bibinfo {author} {\bibfnamefont
			{S.}~\bibnamefont {Wilson}},\ and\ \bibinfo {author} {\bibfnamefont
			{E.}~\bibnamefont {Uykur}},\ }\bibfield  {title} {\bibinfo {title} {Role of
			{Sb} in the superconducting kagome metal {CsV$_3$Sb$_5$} revealed by its
			anisotropic compression},\ }\href
	{https://doi.org/10.21468/SciPostPhys.12.2.049} {\bibfield  {journal}
		{\bibinfo  {journal} {SciPost Physics}\ }\textbf {\bibinfo {volume} {12}},\
		\bibinfo {pages} {49} (\bibinfo {year} {2022})}\BibitemShut {NoStop}%
	\bibitem [{\citenamefont {Yu}\ \emph {et~al.}(2022)\citenamefont {Yu},
		\citenamefont {Zhu}, \citenamefont {Wen}, \citenamefont {Gui}, \citenamefont
		{Li}, \citenamefont {Han}, \citenamefont {Wu}, \citenamefont {Wang},
		\citenamefont {Xiang}, \citenamefont {Qiao}, \citenamefont {Ying},\ and\
		\citenamefont {Chen}}]{RN190}%
	\BibitemOpen
	\bibfield  {author} {\bibinfo {author} {\bibfnamefont {F.}~\bibnamefont
			{Yu}}, \bibinfo {author} {\bibfnamefont {X.}~\bibnamefont {Zhu}}, \bibinfo
		{author} {\bibfnamefont {X.}~\bibnamefont {Wen}}, \bibinfo {author}
		{\bibfnamefont {Z.}~\bibnamefont {Gui}}, \bibinfo {author} {\bibfnamefont
			{Z.}~\bibnamefont {Li}}, \bibinfo {author} {\bibfnamefont {Y.}~\bibnamefont
			{Han}}, \bibinfo {author} {\bibfnamefont {T.}~\bibnamefont {Wu}}, \bibinfo
		{author} {\bibfnamefont {Z.}~\bibnamefont {Wang}}, \bibinfo {author}
		{\bibfnamefont {Z.}~\bibnamefont {Xiang}}, \bibinfo {author} {\bibfnamefont
			{Z.}~\bibnamefont {Qiao}}, \bibinfo {author} {\bibfnamefont {J.}~\bibnamefont
			{Ying}},\ and\ \bibinfo {author} {\bibfnamefont {X.}~\bibnamefont {Chen}},\
	}\bibfield  {title} {\bibinfo {title} {Pressure-induced dimensional crossover
			in a kagome superconductor},\ }\href
	{https://doi.org/10.1103/PhysRevLett.128.077001} {\bibfield  {journal}
		{\bibinfo  {journal} {Phys. Rev. Lett.}\ }\textbf {\bibinfo {volume} {128}},\
		\bibinfo {pages} {077001} (\bibinfo {year} {2022})}\BibitemShut {NoStop}%
	\bibitem [{\citenamefont {Zhang}\ \emph
		{et~al.}(2021{\natexlab{c}})\citenamefont {Zhang}, \citenamefont {Chen},
		\citenamefont {Zhou}, \citenamefont {Yuan}, \citenamefont {Wang},
		\citenamefont {Wang}, \citenamefont {Yang}, \citenamefont {An}, \citenamefont
		{Zhang}, \citenamefont {Zhu}, \citenamefont {Zhou}, \citenamefont {Chen},
		\citenamefont {Zhou},\ and\ \citenamefont {Yang}}]{RN44}%
	\BibitemOpen
	\bibfield  {author} {\bibinfo {author} {\bibfnamefont {Z.}~\bibnamefont
			{Zhang}}, \bibinfo {author} {\bibfnamefont {Z.}~\bibnamefont {Chen}},
		\bibinfo {author} {\bibfnamefont {Y.}~\bibnamefont {Zhou}}, \bibinfo {author}
		{\bibfnamefont {Y.}~\bibnamefont {Yuan}}, \bibinfo {author} {\bibfnamefont
			{S.}~\bibnamefont {Wang}}, \bibinfo {author} {\bibfnamefont {J.}~\bibnamefont
			{Wang}}, \bibinfo {author} {\bibfnamefont {H.}~\bibnamefont {Yang}}, \bibinfo
		{author} {\bibfnamefont {C.}~\bibnamefont {An}}, \bibinfo {author}
		{\bibfnamefont {L.}~\bibnamefont {Zhang}}, \bibinfo {author} {\bibfnamefont
			{X.}~\bibnamefont {Zhu}}, \bibinfo {author} {\bibfnamefont {Y.}~\bibnamefont
			{Zhou}}, \bibinfo {author} {\bibfnamefont {X.}~\bibnamefont {Chen}}, \bibinfo
		{author} {\bibfnamefont {J.}~\bibnamefont {Zhou}},\ and\ \bibinfo {author}
		{\bibfnamefont {Z.}~\bibnamefont {Yang}},\ }\bibfield  {title} {\bibinfo
		{title} {Pressure-induced reemergence of superconductivity in the topological
			kagome metal {{CsV$_{3}$Sb$_{5}$}}},\ }\href
	{https://doi.org/10.1103/PhysRevB.103.224513} {\bibfield  {journal} {\bibinfo
			{journal} {Phys. Rev. B}\ }\textbf {\bibinfo {volume} {103}},\ \bibinfo
		{pages} {224513} (\bibinfo {year} {2021}{\natexlab{c}})}\BibitemShut
	{NoStop}%
	\bibitem [{\citenamefont {Du}\ \emph {et~al.}(2022)\citenamefont {Du},
		\citenamefont {Li}, \citenamefont {Luo}, \citenamefont {Gong}, \citenamefont
		{Li}, \citenamefont {Jiang}, \citenamefont {Ortiz}, \citenamefont {Liu},
		\citenamefont {Xu}, \citenamefont {Wilson}, \citenamefont {Cao},
		\citenamefont {Song},\ and\ \citenamefont {Yuan}}]{RN628}%
	\BibitemOpen
	\bibfield  {author} {\bibinfo {author} {\bibfnamefont {F.}~\bibnamefont
			{Du}}, \bibinfo {author} {\bibfnamefont {R.}~\bibnamefont {Li}}, \bibinfo
		{author} {\bibfnamefont {S.}~\bibnamefont {Luo}}, \bibinfo {author}
		{\bibfnamefont {Y.}~\bibnamefont {Gong}}, \bibinfo {author} {\bibfnamefont
			{Y.}~\bibnamefont {Li}}, \bibinfo {author} {\bibfnamefont {S.}~\bibnamefont
			{Jiang}}, \bibinfo {author} {\bibfnamefont {B.~R.}\ \bibnamefont {Ortiz}},
		\bibinfo {author} {\bibfnamefont {Y.}~\bibnamefont {Liu}}, \bibinfo {author}
		{\bibfnamefont {X.}~\bibnamefont {Xu}}, \bibinfo {author} {\bibfnamefont
			{S.~D.}\ \bibnamefont {Wilson}}, \bibinfo {author} {\bibfnamefont
			{C.}~\bibnamefont {Cao}}, \bibinfo {author} {\bibfnamefont {Y.}~\bibnamefont
			{Song}},\ and\ \bibinfo {author} {\bibfnamefont {H.}~\bibnamefont {Yuan}},\
	}\bibfield  {title} {\bibinfo {title} {Superconductivity modulated by
			structural phase transitions in pressurized vanadium-based kagome metals},\
	}\href {https://doi.org/10.1103/PhysRevB.106.024516} {\bibfield  {journal}
		{\bibinfo  {journal} {Phys. Rev. B}\ }\textbf {\bibinfo {volume} {106}},\
		\bibinfo {pages} {024516} (\bibinfo {year} {2022})}\BibitemShut {NoStop}%
	\bibitem [{\citenamefont {Fu}\ and\ \citenamefont {Kane}(2008)}]{RN440}%
	\BibitemOpen
	\bibfield  {author} {\bibinfo {author} {\bibfnamefont {L.}~\bibnamefont
			{Fu}}\ and\ \bibinfo {author} {\bibfnamefont {C.~L.}\ \bibnamefont {Kane}},\
	}\bibfield  {title} {\bibinfo {title} {Superconducting proximity effect and
			{Majorana} fermions at the surface of a topological insulator},\ }\href
	{https://doi.org/10.1103/PhysRevLett.100.096407} {\bibfield  {journal}
		{\bibinfo  {journal} {Phys. Rev. Lett.}\ }\textbf {\bibinfo {volume} {100}},\
		\bibinfo {pages} {096407} (\bibinfo {year} {2008})}\BibitemShut {NoStop}%
\end{thebibliography}

%

\end{document}